\documentclass[12pt,a4paper]{article}
\pdfoutput=1
\usepackage{jheppub}
\usepackage[sort&compress]{natbib}
\usepackage{url}
\usepackage{hyperref}
\usepackage{amsfonts}% Include figure files
\usepackage{epsfig}\usepackage{dcolumn}% Align table columns on decimal point
\usepackage{bm}% bold math
\usepackage{slashed}
\graphicspath{{./figuras/}}
\usepackage{subfigure}

\newcommand{\zp}{Z^{\prime}}

\newcommand{\mzp}{M_{Z'}}
\newcommand{\mchi}{M_{\chi}}

\newcommand{\gx}{g_{\chi}}

\newcommand{\beq}{\begin{equation}} \newcommand{\eeq}{\end{equation}}
\newcommand{\bea}{\begin{eqnarray}} \newcommand{\eea}{\end{eqnarray}}
           
\newcommand{\pL}{\left(} \newcommand{\pR}{\right)} \newcommand{\bL}{\left[} \newcommand{\bR}{\right]}    
\newcommand{\be}{\begin{equation}}
\newcommand{\ee}{\end{equation}}

\begin{document}
\preprint{LPT-Orsay-16-92}
%\title{{\color{blue}Closing in on the Low Mass Dark $Z^{\prime}$ Portal}}
\title{ {\color{blue}Augury of Darkness: The Low-Mass Dark $Z^{\prime}$ Portal}  }
%\title{The Low Mass Dark $Z^{\prime}$ Portal}

\author{Alexandre Alves$^a$,}
\emailAdd{aalves@unifesp.br}

\author{Giorgio Arcadi$^b$,}
\emailAdd{arcadi@mpi-hd.mpg.de}

\author{Yann Mambrini$^{c}$,}
\emailAdd{yann.mambrini@th.u-psud.fr}

\author{Stefano Profumo$^{d}$,$^{e}$}
\emailAdd{profumo@ucsc.edu}

\author{Farinaldo S. Queiroz$^b$,}
\emailAdd{queiroz@mpi-hd.mpg.de}

\affiliation{$^a$Departamento de F\'isica,
Universidade Federal de S\~ao Paulo, Diadema-SP, 09972-270, Brasil}
%\affiliation{$^{b}$High Energy Physics Division, Argonne National Laboratory, Argonne, IL 60439, USA}
\affiliation{$^b$Max-Planck-Institut fur Kernphysik, Saupfercheckweg 1, 69117 Heidelberg, Germany}
%\affiliation{$^{e}$ Enrico Fermi Institute, University of Chicago, Chicago, IL 60637, USA}
%\affiliation{$^{f}$ Kavli Institute for Cosmological Physics, University of Chicago, Chicago, IL 60637, USA}
\affiliation{$^c$Laboratoire de Physique Th\'eorique, CNRS -- UMR 8627,
Universit\'e de Paris-Saclay 11, F-91405 Orsay Cedex, France}
\affiliation{$^d$Department of Physics, University of California, Santa Cruz, 1156 High St, Santa Cruz, CA 95060, United States of America}
\affiliation{$^e$Santa Cruz Institute for Particle Physics, Santa Cruz, 1156 High St, Santa Cruz, CA 95060, United States of America}

\abstract{
Dirac fermion dark matter models with heavy $Z^{\prime}$ mediators are subject to stringent constraints from spin-independent direct searches and from LHC bounds, cornering them to live near the $Z^{\prime}$ resonance. Such constraints can be relaxed, however, by turning off the vector coupling to Standard Model fermions, thus weakening direct detection bounds, or by resorting to light $Z^{\prime}$ masses, below the Z pole, to escape heavy resonance searches at the LHC. In this work we investigate both cases, as well as the applicability of our findings to Majorana dark matter. We derive collider bounds for light $Z^{\prime}$ gauge bosons using the $CL_S$ method, spin-dependent scattering limits, as well as the spin-independent scattering rate arising from the evolution of couplings between the energy scale of the mediator mass and the nuclear energy scale, and indirect detection limits. We show that such scenarios are still rather constrained by data, and that near resonance they could accommodate the gamma-ray GeV excess in the Galactic center.}

%apply, whereas for vector-axial interactions less stringent spin-dependent limits structure. Thus the nature of the dark matter particle determines the dark matter phenomenology. Although models, we call for short "Twisted", can give rise to null vectorial couplings between the gauge boson rising from a $U(1)_X$ symmetry and the SM fermions simply by the gauge coupling assignments. Thus, a Dirac fermion dark matter particle mediated by a $Z^{\prime}$ boson, has dominantly spin-dependent scatterings, conversely of the canonical case, hence ``twisted''. In this context we study the dark matter phenomenology considering, direct detection limits from PICO and ICECUBE, and indirect detection from CMB and Dwarf Spheroidal galaxies. We probe both heavy and light $Z^{\prime}$ mass regimes, with special attention to the latter, using dimuon data from LHC and derive limits stemming from the  muon magnetic moment. Lastly, we show that this model can accommodate the GeV gamma-ray excess in the Galactic Center also known as ``Hooperon''.

%\pacs{95.35.+d, 14.60.Pq, 98.80.Cq, 12.60.Fr}
\maketitle
\flushbottom

\section{Introduction}

Non-baryonic dark matter (DM) accounts for about 27\%  of the energy budget of the universe  \cite{Ade:2015xua}. Its particle nature is one of the most pressing puzzles at the interface of particle physics and cosmology. Several dark matter candidates have been extensively discussed and reviewed in the literature (see e.g. \cite{Feng:2010gw,Bertone:2010zza}); among those, Weakly Interacting Massive Particles (WIMPs) stand out for arising in several compelling particle physics models, such as supersymmetry,  for naturally accounting for the DM abundance in the universe through the thermal freeze-out paradigm, and for potentially being testable with current and future experimental probes (see e.g. \cite{Baer:2014eja,Strigari:2013iaa,Bertone:2004pz}). 

The key strategies for WIMP searches are direct, indirect, and collider searches. The former consist of measuring  nuclear scattering events with recoil energies on the order of the keV in underground laboratories \cite{Cremonesi:2013bma,Undagoitia:2015gya,Mayet:2016zxu,Queiroz:2016sxf}. WIMP signals in a direct detection experiment are directly proportional to the local dark matter density, thus the observation of a signal can be strongly tied to the presence of WIMP scattering. 

Indirect detection attempts to detect the stable Standard Model particle products of dark matter annihilation, such as gamma-rays, cosmic-rays, or radiation at lower frequency in the electromagnetic spectrum \cite{Profumo:2013yn,Conrad:2014tla,Queiroz:2016awc,Fornengo:2011iq,Mambrini:2012ue}. The signal observed is proportional to the integrated line-of-sight dark matter density squared in the region of interest. 

Finally, collider searches hinge on the fact that high-energy proton-proton collisions at the LHC can generate dark matter particles in association with other exotic particles. The associated signature would consist of missing energy in, for instance, monojet or dijet searches. Whilst not  capable to unveil the astrophysical connection of the particles produced, collider studies can provide a complementary and sometimes more effective way to constrain dark matter models \cite{Abercrombie:2015wmb},
especially with light dark matter particles. 

The efficacy of each detection strategy at probing WIMPs is rather model-dependent; however, and rather interestingly, for the model we focus on this paper, there is a remarkable degree of complementarity across direct, indirect, and collider searches.

The observation of WIMP events at any of the detection strategies would be paramount to understand the laws of nature at fundamental scales, since WIMPs are expected to be embedded in UV complete models such as the minimal superymmetric standard models or minimal left-right model \cite{Edsjo:1997bg,Griest:1988ma,Profumo:2005xd,Garcia-Cely:2015quu}. In other words, the discovery of WIMPs is tightly related to uncovering hints about underlying physics beyond the Standard Model. 

In order to map the interactions between WIMPs and standard model particles which are allowed by data, simplified models have become powerful tools. In particular, simplified models which make use of vector mediators \cite{Karwin:2016tsw}.

%Although, since not much is known regarding the WIMP properties, they may interact with SM model particles in many ways, 
%but some are rather testable, as when a neutral gauge bosons $Z'$ act as portal between the SM and dark matter particles. In order to map the possible interactions allowed by data, simplified models have become powerful tools. In this endevour the coupling constants are kept free to some extent, but keep in mind that in a UV complete model, typically the coupling strength as well as the interactions are determined by the underlying gauge group %{\color{red} Yann: what do you mean by that? In unified models you mean? In what sense the coupling strength is determined by the gauge group?}. 

Models with a $Z^\prime$ neutral gauge boson portal between dark and ordinary matter have attracted significant attention for a variety of reasons: they for instance represent ``simplified model'' version of several compelling particle models, and are constrained by data in a rather stringent way, albeit the couplings of the new boson to dark and ordinary matter are largely model-dependent \cite{Mizukoshi:2010ky,Cogollo:2012ek,Alvares:2012qv,Cogollo:2014jia,Martinez:2014ova,Kelso:2014qka,Alves:2014yha,Allanach:2015gkd,Dong:2014esa,Dong:2015rka,Alves:2016fqe}.

Assuming the dark matter particle to be a Dirac fermion, many analysis have been done in the context of heavy mediators ($M_{Z^{\prime}} > 1$~TeV) 
\cite{Gondolo:2011eq,An:2012ue,Profumo:2013sca,Alves:2013tqa,
Arcadi:2013qia,deSimone:2014pda,Cline:2014dwa,Chen:2014noa,Feng:2014eja,Lebedev:2014bba,Dong:2014wsa,Kahlhoefer:2015bea,Marcos:2015dza,
Chala:2015ama,Chen:2015tia,Ducu:2015fda,Martin-Lozano:2015vva,Chiang:2015ika,Okada:2016tci,Celis:2016ayl,Klasen:2016qux,
Bell:2016uhg,Ko:2016ala,Duerr:2016tmh,Fairbairn:2016iuf,
Jacques:2016dqz,Englert:2016joy,Sage:2016uxt,Altmannshofer:2016jzy,Okada:2016gsh}. The key results are that these models are plagued with restrictive spin-independent direct detection limits as well as LHC bounds on the $Z^{\prime}$ mass from heavy resonance searches, limiting the allowed parameter space to the $Z^{\prime}$ resonance, i.e. when the mass of the dark matter is close to half the mass of the $Z^{\prime}$. 

In this work, we investigate an alternative scenario by turning off the vector coupling to Standard Model fermions as proposed in \cite{Lebedev:2014bba} to weaken direct detection bounds, and by focusing on relatively light $Z^{\prime}$ masses, ($M_{Z^{\prime}} < 500$~GeV) , to circumvent the usual heavy-resonance searches at the LHC \footnote{See also Ref.\cite{Bai:2015nfa} for an study on light $Z^{\prime}$ bosons, focused on mono $Z^{\prime}$  signatures at the LHC.}.

The present analysis markedly differs from previous analysis for a variety of reasons:

{\color{blue} (i)} We focus on a very specific class of $Z^{\prime}$ models, namely those where the $Z^{\prime}$ possesses purely axial-vector couplings with SM fermions, and we perform a detailed dark matter phenomenology study;\\

{\color{blue} (ii)} We show that the $Z^{\prime}$ mass can be as low as $15$~GeV, where the heavy resonance searches at the LHC searches are not applicable. We explicitly compute the collider limits in that region, with no rescaling, using the $CL_S$ method employing dimuon data from the LHC; \\

{\color{blue} (iii)} We discuss the possibility of accommodating the gamma-ray excess observed in the Galactic center in the context of this class of models. \\

\noindent
The paper is structured as follows. We introduce the model under consideration in Section 2. Section 3 is devoted to a detailed study of the invisible $Z'$ searches at LHC, whereas
direct detection constraints are analyzed in Section 4. After a discussion on 
the Galactic center excess in Section 6, we conclude.

%We begin our reasoning by introducing the $Z^{\prime}$ model under scrutinity and %how it can be realized in a UV complete framework. 
% study with an outline of the $\zp$ DM portal we focus on.

\section{Model}

We investigate here a $U(1)_X$ extension of the Standard Model expected to be less constrained by collider, direct and indirect detection searches. The model is based on the gauge group $SU(3)_c \otimes SU(2)_L \otimes U(1)_Y \otimes U(1)_X$. Augmenting the SM by a new Abelian symmetry implies the existence of a new gauge boson $Z'$, which can gain mass in different ways. To preserve gauge invariance such gauge boson will couple to SM fermion through the covariant derivatives $\bar{f}_L\gamma_{\mu}D^{\mu}f_L$ and $\bar{f}_R\gamma_{\mu}D^{\mu}f_R$, where $D^{\mu}=\partial^{\mu}-i\, g_f\, q_f Z^{\prime \mu}$, which lead to,
%\color{red} the "i" is missing in the following formula

\begin{equation}
\mathcal{L}  \supset i \bar{f}\gamma_{\mu} \left[\partial_\mu - i g_f \frac{q_{f\,L}+q_{f\,R}}{2}-i g_f \frac{q_{fR}-q_{fL}}{2} \gamma^5 \right] f\, Z^{\prime \mu} 
\end{equation}

If $q_{fL}=q_{fR}$, i.e. the left and right-handed SM fermions transform in the same way under $U(1)_X$ (vector-like fermions), the $Z^{\prime}$ will have only vectorial couplings with SM fermions, corresponding to a dark photon. 
Conversely, if $q_{fL}=-q_{fR}$, only axial-vector current are non-vanishing. The latter is the scenario we are interested in. The addition of a Dirac fermion dark matter field is trivial and follows the same logic. 
%{\color{red} YANN: stupid question, if the SM fermions are axial + vector but the DM is only axial, 
%does the result change a lot? The mixed terms vector(SM)-axial(DM) should vanish, no? For instance a Majorana DM should have only axial coupling to $Z'_mu$, no?
%So whatever is the nature of the coupling to SM, it is axial-axial which will stay at the end. Surely I am missing something..}
Focusing on the latter the final Lagrangian reads

\begin{equation}
\label{eq:Diracfermion}
\mathcal{L} \supset \left[ \bar{\chi} \gamma^\mu ( g_{\chi  v } + g_{\chi a} \gamma^5 ) \chi + g_f \bar{f} \gamma^\mu \gamma^5  f \right] Z'_\mu, \\
\end{equation} where $\chi$ is the dark matter candidate %\color{red} Should we insist on the Dirac or Majorana nature of the DM? For instance, a Majorana would have $g_v=0$} and we took $q_{fL}=q_{fR}=1$ without loss of generalities.

We remark that in order to write a Lagrangian of the form Eq. (\ref{eq:Diracfermion}) it is necessary to assume that SM fermions be charged under the $U(1)_X$ symmetry. 
%the $Z^{\prime}$ gauge boson stems from. 
One should also notice that the model is clearly anomalous: due
to the chirality of the SM fermions, the triangle anomalies $U(1)_X^3$
do not cancel.  Anomaly cancellation generically requires the existence of new fields. The new fields can, however, be vector-like under the SM gauge group, while being chiral under the new Abelian symmetry. With appropriate charge assignments one can construct an anomaly-free model where the $Z^{\prime}$ has only axial-vector coupling to fermions. In Ref.~\cite{Ismail:2016tod}, the authors have put some effort in coming up with UV complete models where the Eq. (\ref{eq:Diracfermion}) is realized. We will thus assume that the exotic fermions needed to cancel the anomalies are sufficiently heavy so as not to spoil the dark matter phenomenology\footnote{This is not always possible, as argued in 
\cite{Ismail:2016tod}, since the exotic fermions may contribute to the renormalization group equation and affect the running of the couplings.}. 
We emphasize that this assumption is crucial to the validity of our results, especially because we will be focusing on $Z^{\prime}$ masses below 1 TeV.

All the numerical computations will be carried under the assumption $g_{\chi  v }=g_{\chi  a }=g_\chi$. Keeping them in the same order is arguably a natural choice. 
Mild departures from this assumption will change neither the relic density nor the annihilation cross section today since they are both dominated by the vectorial term. As for WIMP-nucleon scattering rates, the impact is also mild. However, had we set $g_{\chi  v }$ to zero, we would have been discussing a Majorana fermion, where the annihilation cross section is helicity suppressed, and the WIMP-nucleon scattering is purely spin-dependent. An overall minus sign between the couplings will induce no change to our results. Furthermore, this choice conveniently reduces the number of free parameters of the simplified model. That said, as long as one does not dramatically deviates from $g_{\chi  v }\, \sim  \, g_{\chi  a }$, our conclusions will readily apply.

%Now we have discussed the theoretical aspects of this simplified model

\section{Collider Constraints on Light $Z^{\prime}$ Models}
Searches for high- and low-mass dilepton resonances at the LHC have been an excellent probe of models containing new neutral vector bosons~\cite{Chatrchyan:2013tia,Aad:2014cka}. In the case where the new vector boson mediates the interaction between the SM and the dark sector, constraints from  dijets and monojet searches for the $\zp$ are complementary in the {\it mass versus coupling} plane~\cite{Alves:2013tqa}. These are the most stringent constraints for leptophobic dark $\zp$ models. When  couplings to leptons are sizable, though, dileptons searches have the potential to exclude larger portions of the models' parameter space~\cite{Alves:2015pea,Alves:2015mua,Hoenig:2014dsa} compared do dijets. This can be understood in view of the relative size of the production cross section for dijets and dileptons and their correspondent irreducible backgrounds: First, both production mechanisms are electroweak processes; second the dominant backgrounds for dijets and dileptons are the QCD jet pair production and the Drell-Yan processes, respectively. For universal fermion couplings as those assumed in this work, the relative number of flavors and color multiplicity leads to the relation (at LO) $\sigma(pp\to\zp\to jj)/\sigma(pp\to\zp\to \ell^+\ell^-)=15$, where $\ell$   denotes electrons or muons. On the other hand, at LO, for the dominant backgrounds we have $\sigma(pp\to Z\to \ell^+\ell^-)/\sigma(pp\to jj)\sim {\cal O}(10^{-4})$ at the 13 TeV LHC~\cite{Alwall:2014hca}, and a similar ratio should be expected at 7 and 8 TeV center-of-mass energies.

\subsection{Signal simulation and branching ratios}
In order to evaluate the constraints from the 7 TeV LHC data~\cite{Chatrchyan:2013tia} below the $Z$ pole, and above it with 8 TeV data~\cite{Aad:2014cka}, we implemented the axial $\zp$ model in \texttt{FeynRules}~\cite{Alloul:2013fw} to simulate our signal events. We also obtained the partial widths for the $\zp$ decays to leptons, jets, dark matter pairs and top pairs. The branching ratios and cross sections depend on four basic parameters: $\{\mzp, \mchi, \gx, g_f\}$, the mass of the $\zp$, the dark matter mass, the $\zp$ coupling to $\chi$, and the (axial) $\zp$ coupling to the SM fermions, respectively.

\begin{figure}[!t]
\centering
\includegraphics[scale=0.7]{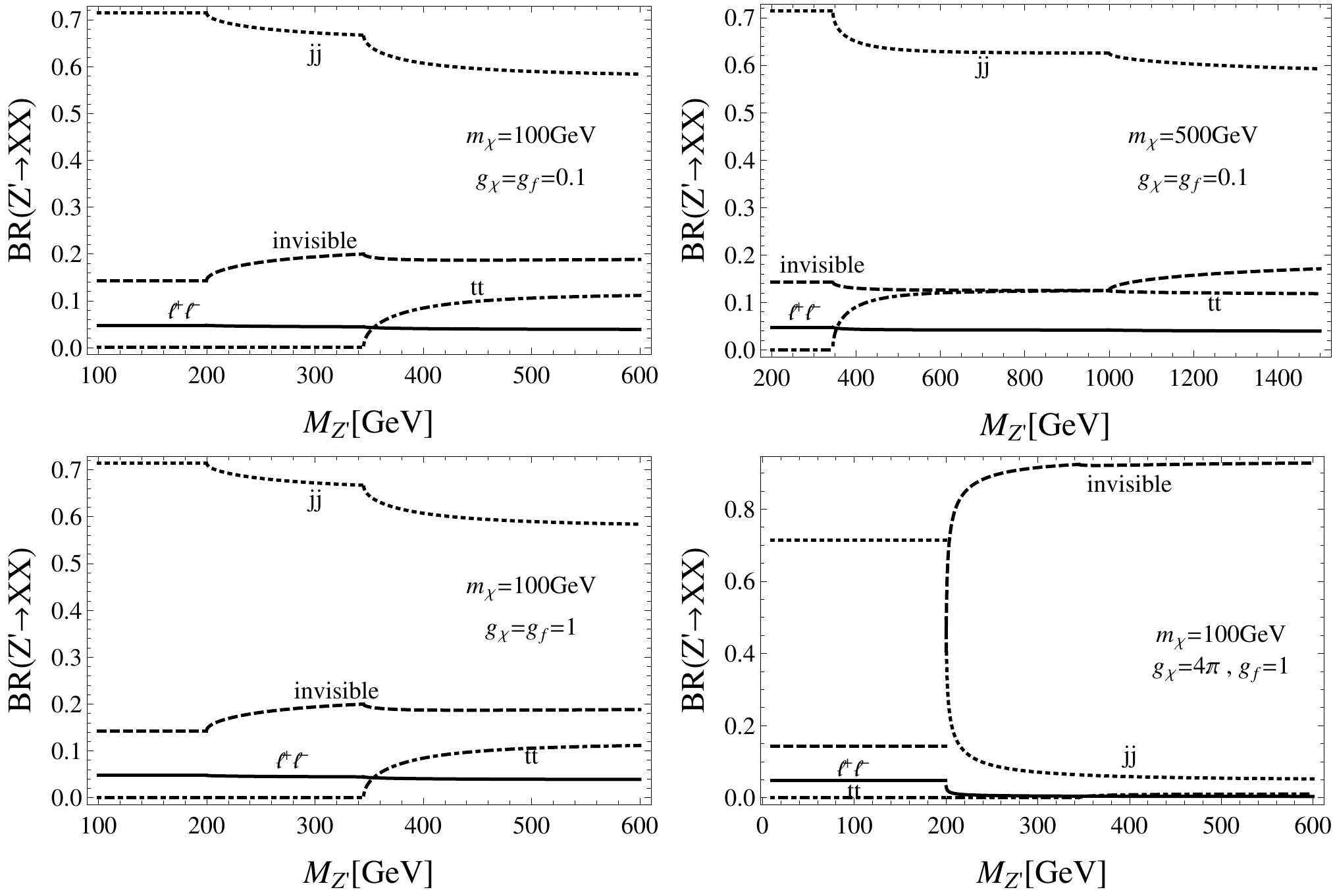}
\caption{The $\zp$ branching ratios to jets, leptons (electrons, muons or taus), top quarks and invisible (DM and neutrinos), as a function of its mass $M_{Z'}$. We present four scenarios: at the left column we fix $m_\chi=100$ GeV, in the upper(lower) panel the couplings are chosen as $g_\chi=g_f=0.1$(1); at the right upper panel we choose a heavier DM with $m_\chi=500$ GeV and $g_\chi=g_f=0.1$, and in the right lower panel we show the branching ratios for an 100 GeV DM, $g_f=1$ and $g_\chi=4\pi$ at the boundary of the perturbative regime.}
\label{brs}
\end{figure}

In Fig.~(\ref{brs}) we show the $Z'$ branching ratios as function of its mass for some benchmark points. 
In the upper left panel we fixed $\mchi=100$ GeV and $\gx=g_f=0.1$. We see that decays to jets dominate, followed by invisible decays, from light to heavy $\zp$ masses, while the branching ratio to leptons (electrons or muons) is of order 3\%. We also observe thresholds when the vector boson is heavy enough to decay to $\chi$ and top pairs. The picture is essentially the same as either $\chi$ gets heavier or the couplings are changed but kept equal to each other, as shown at the upper right panel and the lower left panel. However, the branching ratio to dark matter reaches almost 90\% when $\gx \gg g_f$. In this regime it is possible that a monojet search becomes as competitive as the dileptons concerning the exclusion constraints from collider data.

\begin{figure}[!t]
\centering
\includegraphics[scale=0.41]{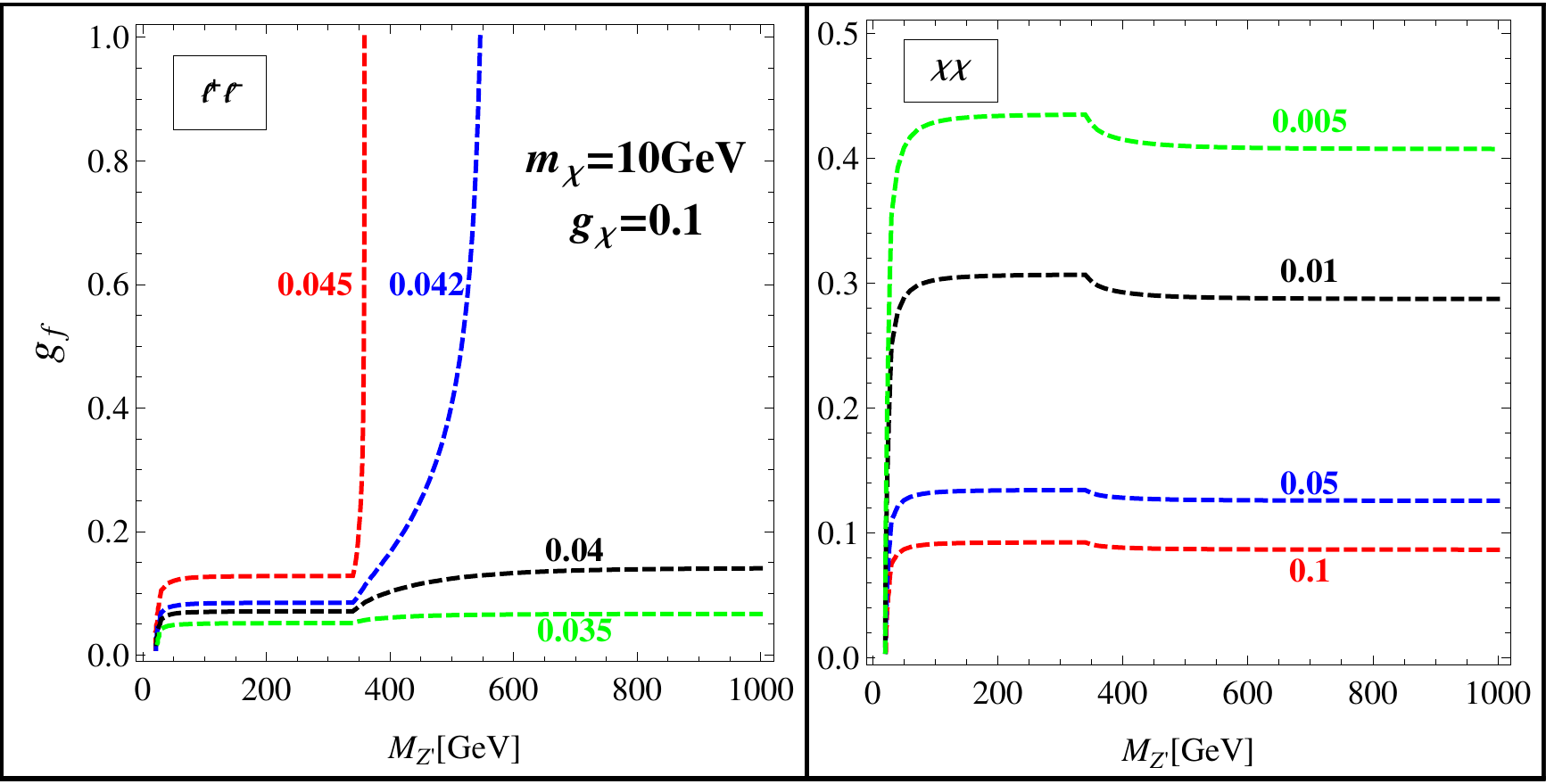}
\includegraphics[scale=0.41]{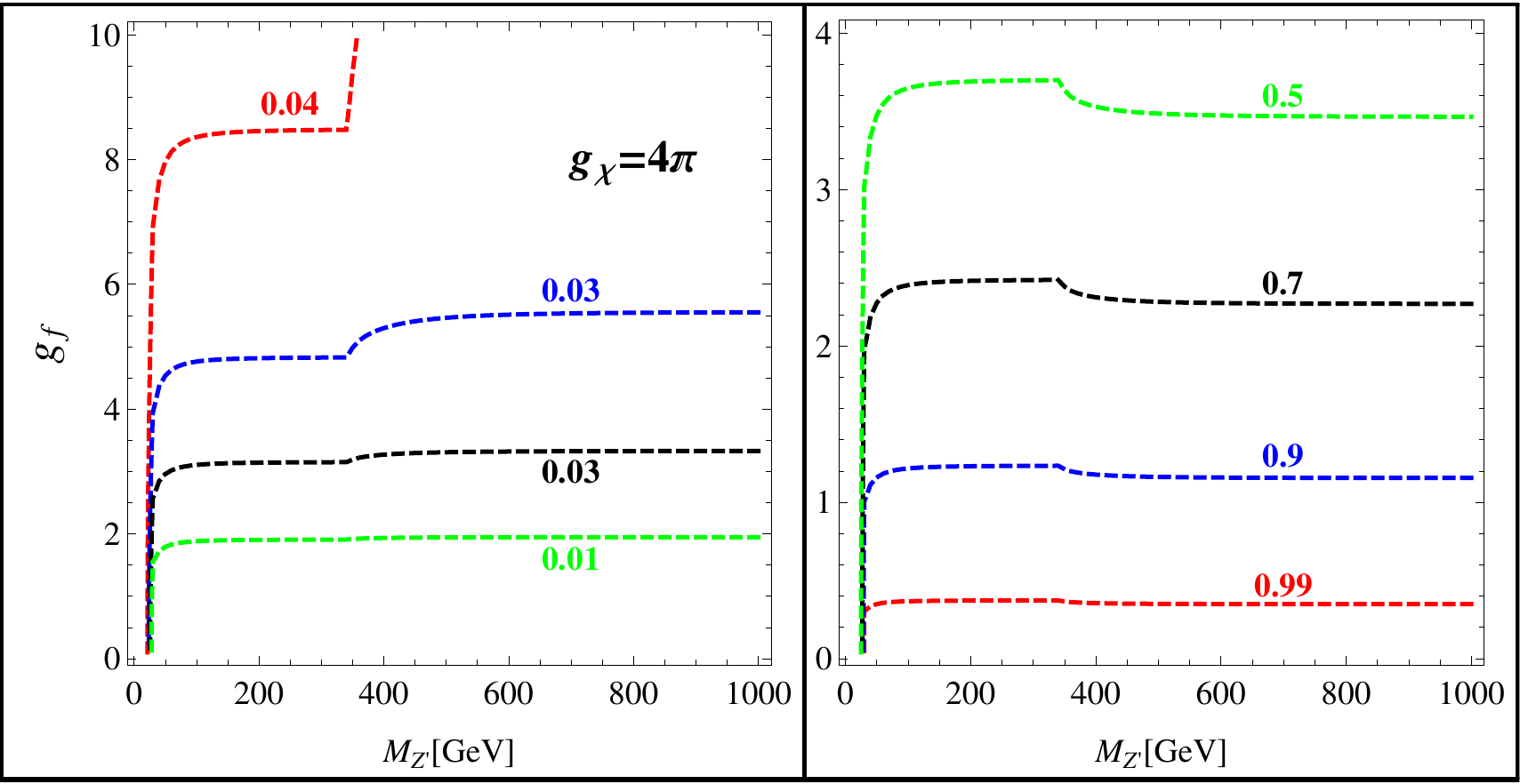}\\
\includegraphics[scale=0.41]{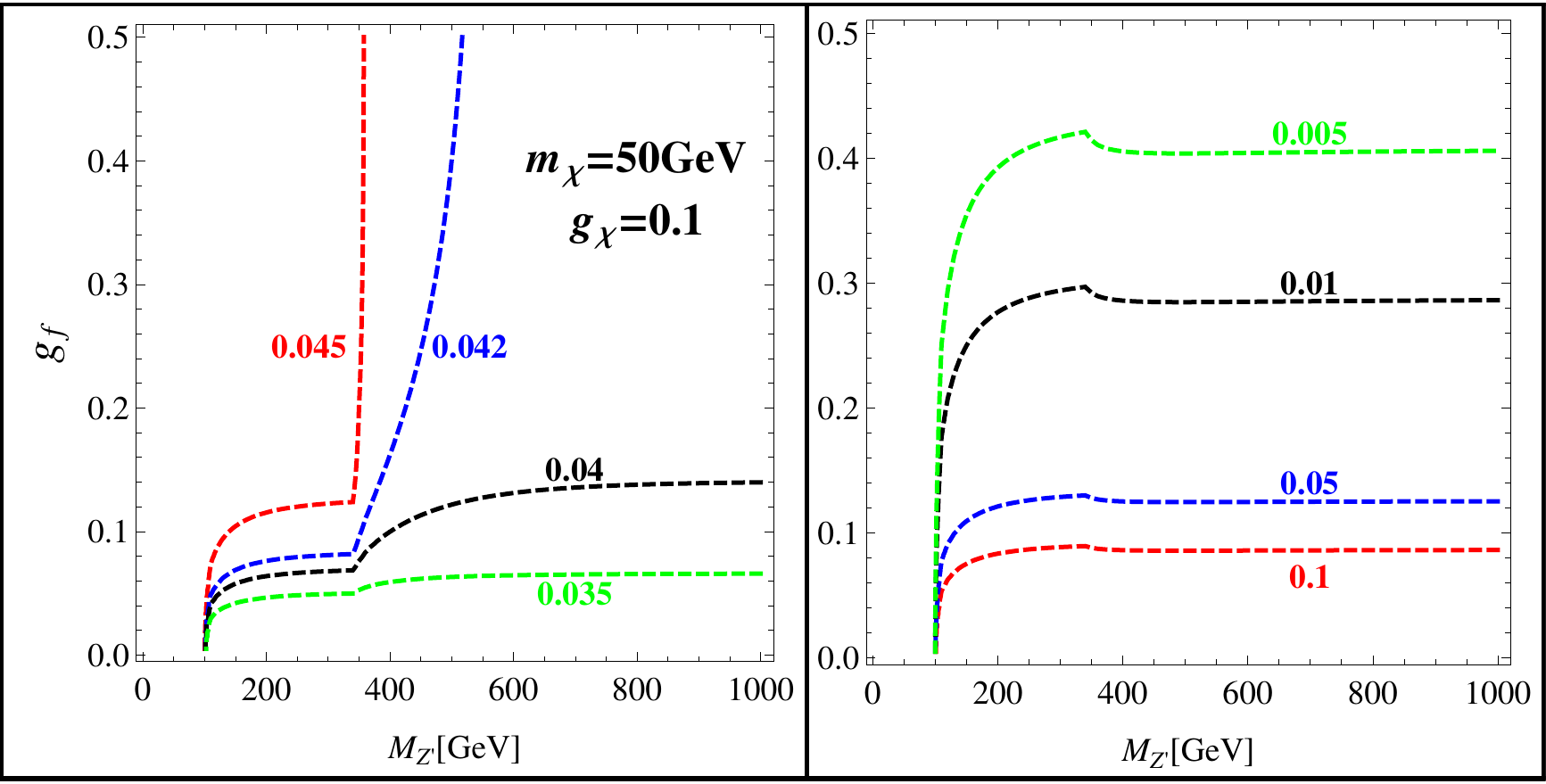}
\includegraphics[scale=0.41]{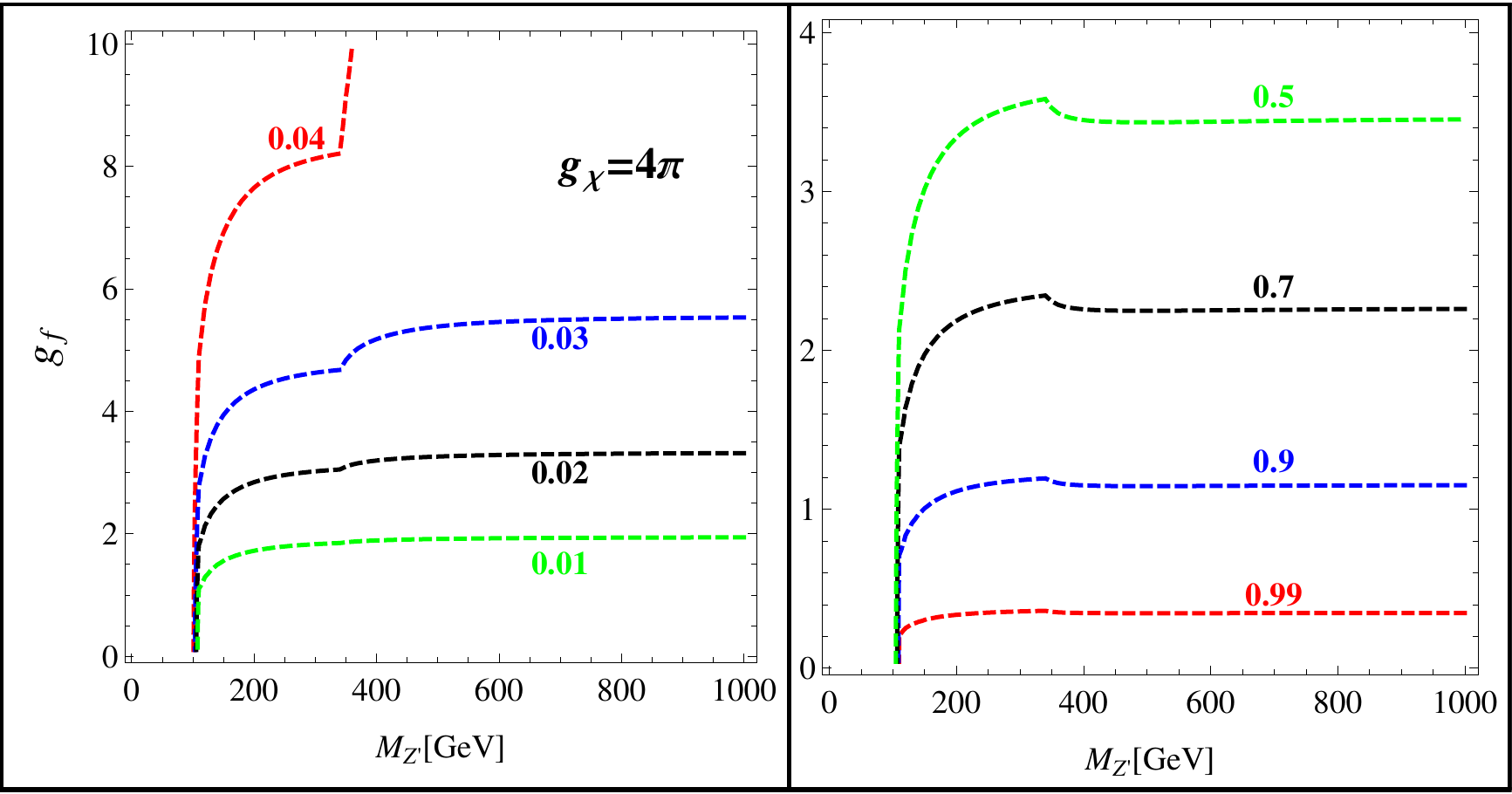}\\
\includegraphics[scale=0.415]{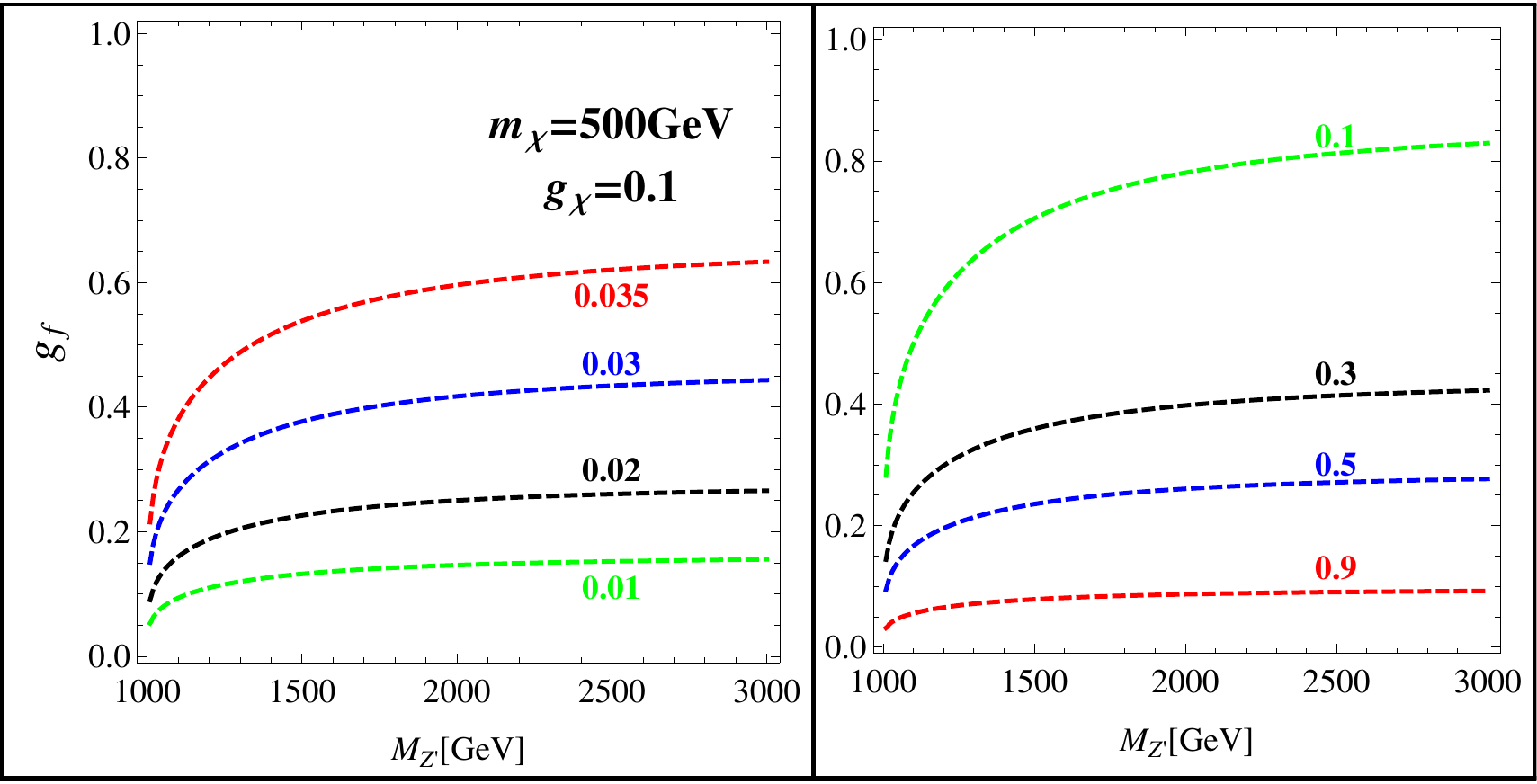}
\includegraphics[scale=0.415]{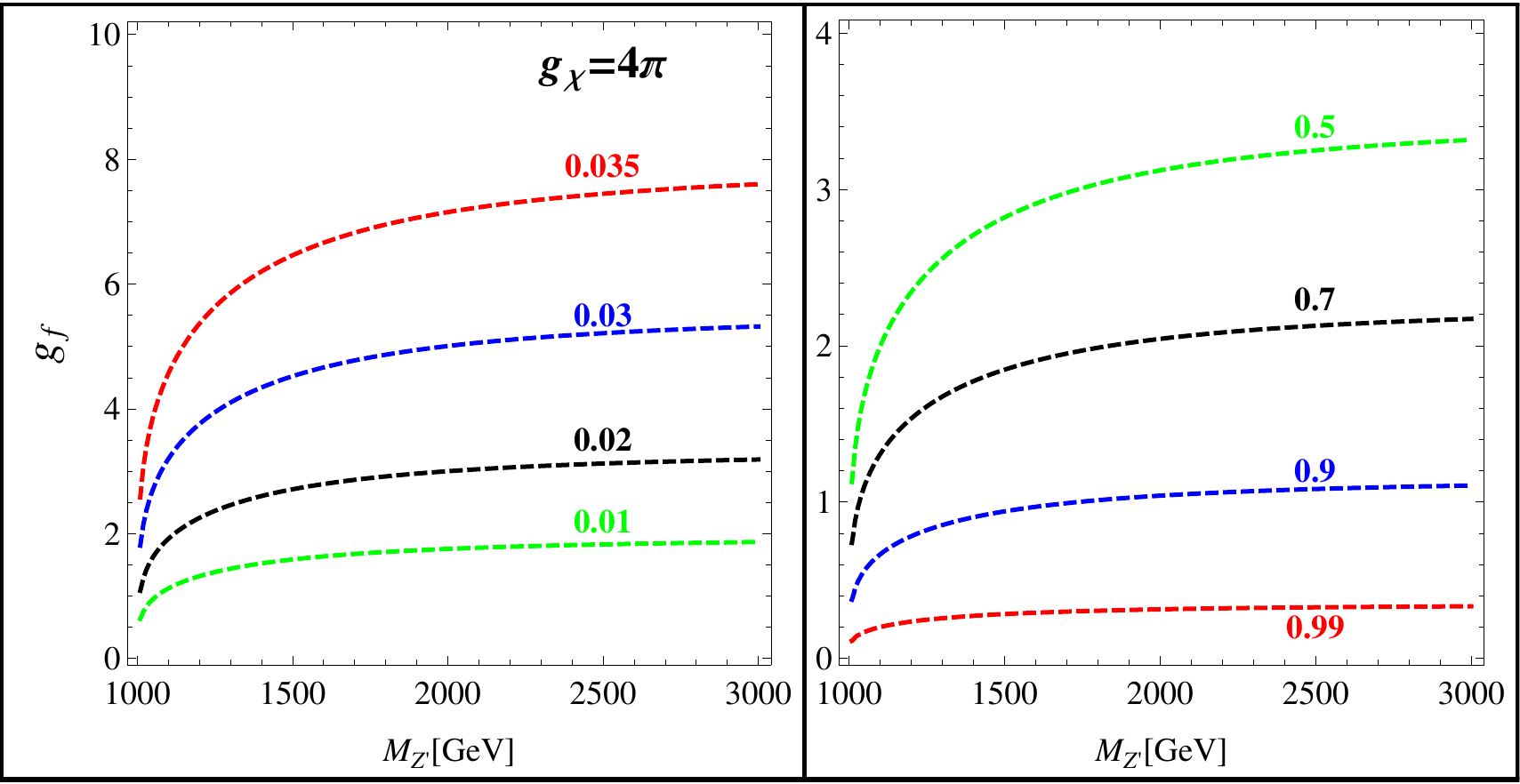}
\caption{The branching ratios into leptons (electrons, muons or taus) and DM in the $\mzp$ {\it versus} $g_f$ plane. At each pair of panels, the left one displays the branching to leptons, and the right one to dark matter. In the first, middle, and last rows we fixed $m_\chi=10$, 50, and 500 GeV, respectively. The left column of plots have $g_\chi=0.1$, while the at right column $g_\chi=4\pi$. The dashed lines represent fixed branching ratios in the mass--coupling plane.}
\label{brgfmz}
\end{figure}

In the $g_f$ {\it versus} $\mzp$ plane, the branching ratio to leptons (muons or electrons) and to invisible (DM plus neutrinos) are shown in the Fig.~(\ref{brgfmz}). In the upper, middle, and lower rows we display the branching ratios for $\mchi=10$, 50, and 500 GeV, respectively. In the left(right) column we fixed $\gx=0.1(4\pi)$. The panels are split into two sub-panels: at left, the branching to leptons, and at right, to invisible.

In the weak DM--$\zp$ coupling regime ($\gx=0.1$) and lighter DM masses ($\mchi \le 50$ GeV), the branching ratio to electrons or muons reaches 4.5\% for all $g_f$ until the top channel opens. The DM decays are low for all $g_f$ as can be seen at the right subpanels. In these scenarios, the dijet channel is the dominant one. As the DM masses increases, a heavy $\zp$ decays mainly to DM as $g_f$ gets small, reaching a 90\% rate for $g_f\sim 0.1$. At the limit of the perturbative regime ($\gx=4\pi$), a $\zp$ decays to DM predominantly, unless $g_f\gtrsim 0.4$. The branching ratio to leptons is considerably suppressed in these scenarios, being at the 1\% level for $g_f\sim 1$ as we see in the right column of Fig.~(\ref{brgfmz}).

\subsection{Searches for dimuon resonances at the 7 and 8 TeV LHC}
Searches for dileptons pairs with invariant masses as low as 15 GeV have been performed by the CMS collaboration~\cite{Chatrchyan:2013tia} at the 7 TeV run with 4.5 fb$^{-1}$. Higher invariant masses up to 4.5 TeV were probed at the 8 TeV LHC by ATLAS with $\sim 20$ fb$^{-1}$~\cite{Aad:2014cka}, for example, both in the dielectron as in the dimuon channel.

We use the low and high mass dimuons from the CMS and ATLAS results, respectively, in order to investigate the collider constraints on the model. Signals for muon pair production were generated with \texttt{MadGraph}~\cite{Alwall:2011uj} with one extra QCD jet, and then interfaced with \texttt{Pythia}~\cite{Sjostrand:2006za} for showering and hadronization simulations. Detector effects and jet clustering were taken into account with \texttt{Delphes}~\cite{deFavereau:2013fsa}. Jet matching were performed in the MLM scheme~\cite{Mangano:2006rw}. The backgrounds, as the data, were taken from the experimental studies~\cite{Chatrchyan:2013tia,Aad:2014cka}.

The dimuons pairs were selected according to the following criteria:
\vskip0.5cm
\noindent \underline{Low mass region}

In the $15<M_{\ell\ell}<100$ GeV invariant mass region, CMS 7 TeV~\cite{Chatrchyan:2013tia} adopted very loose criteria to select dimuon pairs:
\begin{equation}
p_T(\mu_1) > 14\; \hbox{GeV}\;\; ,\;\; p_T(\mu_2) > 9\; \hbox{GeV}\;\; ,\;\; |\eta_\mu| < 2.4
\label{cut:cms7}
\end{equation} 

\noindent \underline{High mass region}

To search for high mass resonances, $M_{\ell\ell}>100$ GeV, with muon pairs, ATLAS 8 TeV~\cite{Aad:2014cka} impose somewhat tighter cuts
\begin{equation}
p_T(\mu_1) > 25\; \hbox{GeV}\;\; ,\;\; p_T(\mu_2) > 25\; \hbox{GeV}\;\; ,\;\; |\eta_\mu| < 2.47
\label{cut:atlas8}
\end{equation} 
Moreover, the muons are required to be isolated. We adopted the same isolation criteria of the experimental collaborations in the \texttt{Delphes} settings. Be aware the slightly stronger limits are currently available from the LHC run-II with $13$~TeV using $13.3\, {\rm fb^{-1}}$ of data for $m_{Z^{\prime}} > 500$~GeV \cite{ATLAS2}. We estimate these limits to be stronger by a factor of 1.3 on the $Z^{\prime}$ mass. Since our focus is on light $Z^{\prime}$ gauge bosons, and our conclusions do not change even with the inclusion of more recent data, we simply keep this older data set.

\subsection{Statistical analysis and estimated bounds}
To estimate the bounds imposed on the $\zp$ masses and couplings we compared the dimuon invariant mass distributions of signal, background and data in the low and high mass regions with
\begin{equation}
\chi^2(\mu_s)=\min_{\{\mu_b\}}\sum_{i}\frac{(d_i-\mu_s s_i-\mu_b b_i)^2}{\mu_s s_i+\mu_b b_i}
\end{equation}
where $d_i$, $b_i$ and $s_i$ represent the $i$-th bin count of the $M_{\ell\ell}$ distribution for data, background and signal, respectively. Our model have two free parameters: $\mu_s$ for signal and $\mu_b$ for the background normalization. The $\mu_b$ parameter is set to the best value that fits the data for a given $\mu_s$.

\begin{figure}[!t]
\centering
\includegraphics[scale=0.4]{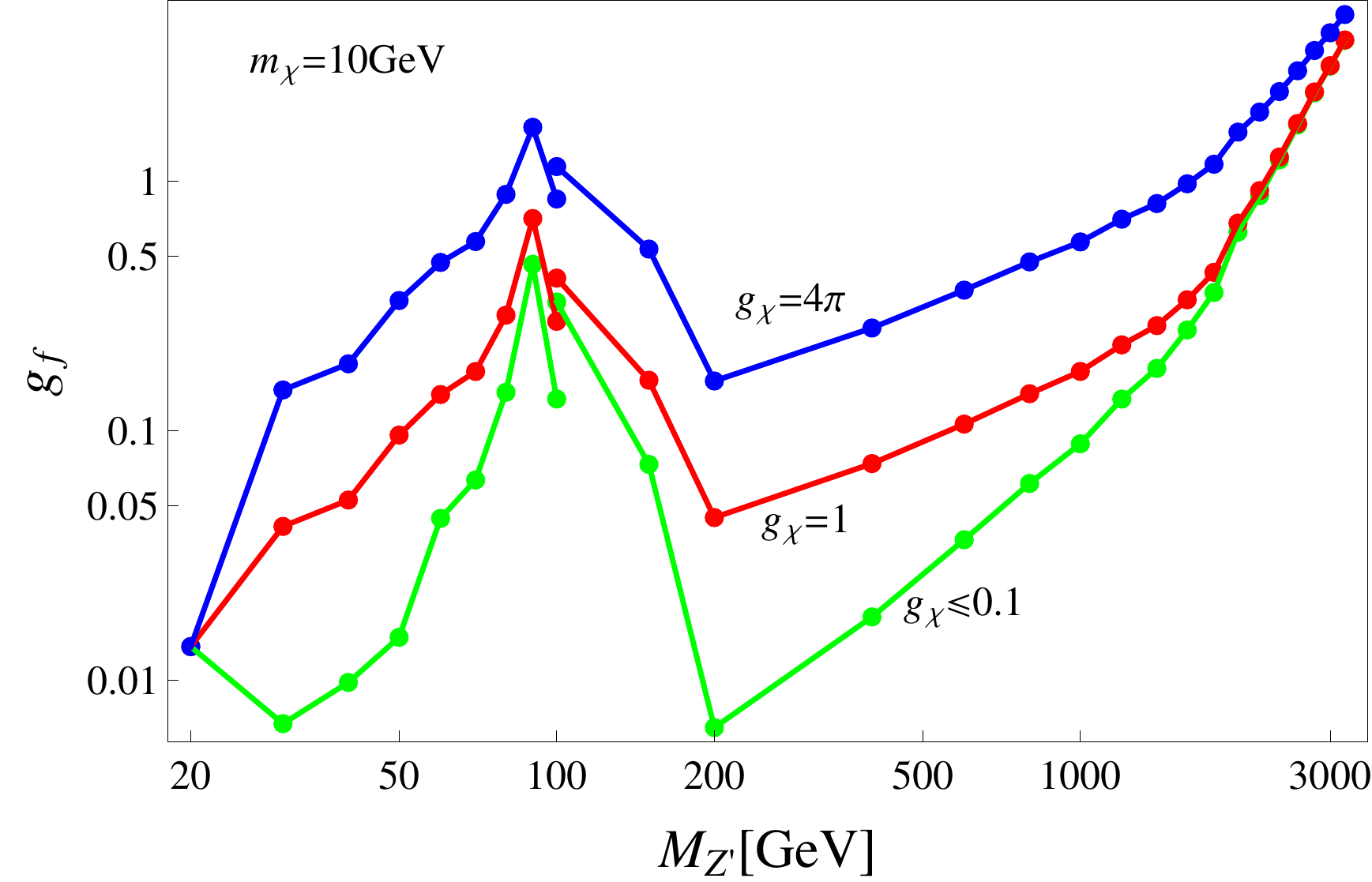}
\includegraphics[scale=0.4]{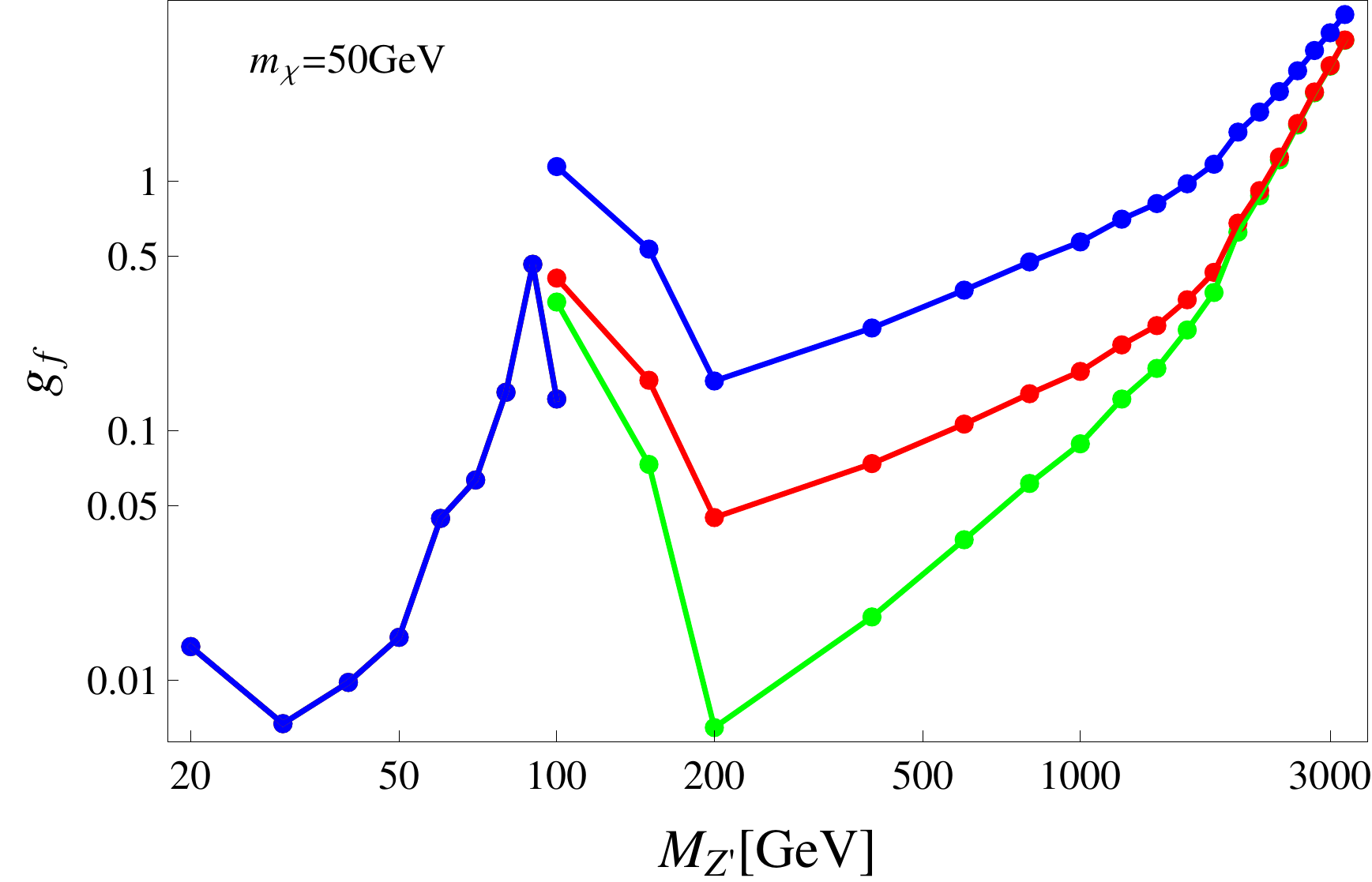}\\
\includegraphics[scale=0.4]{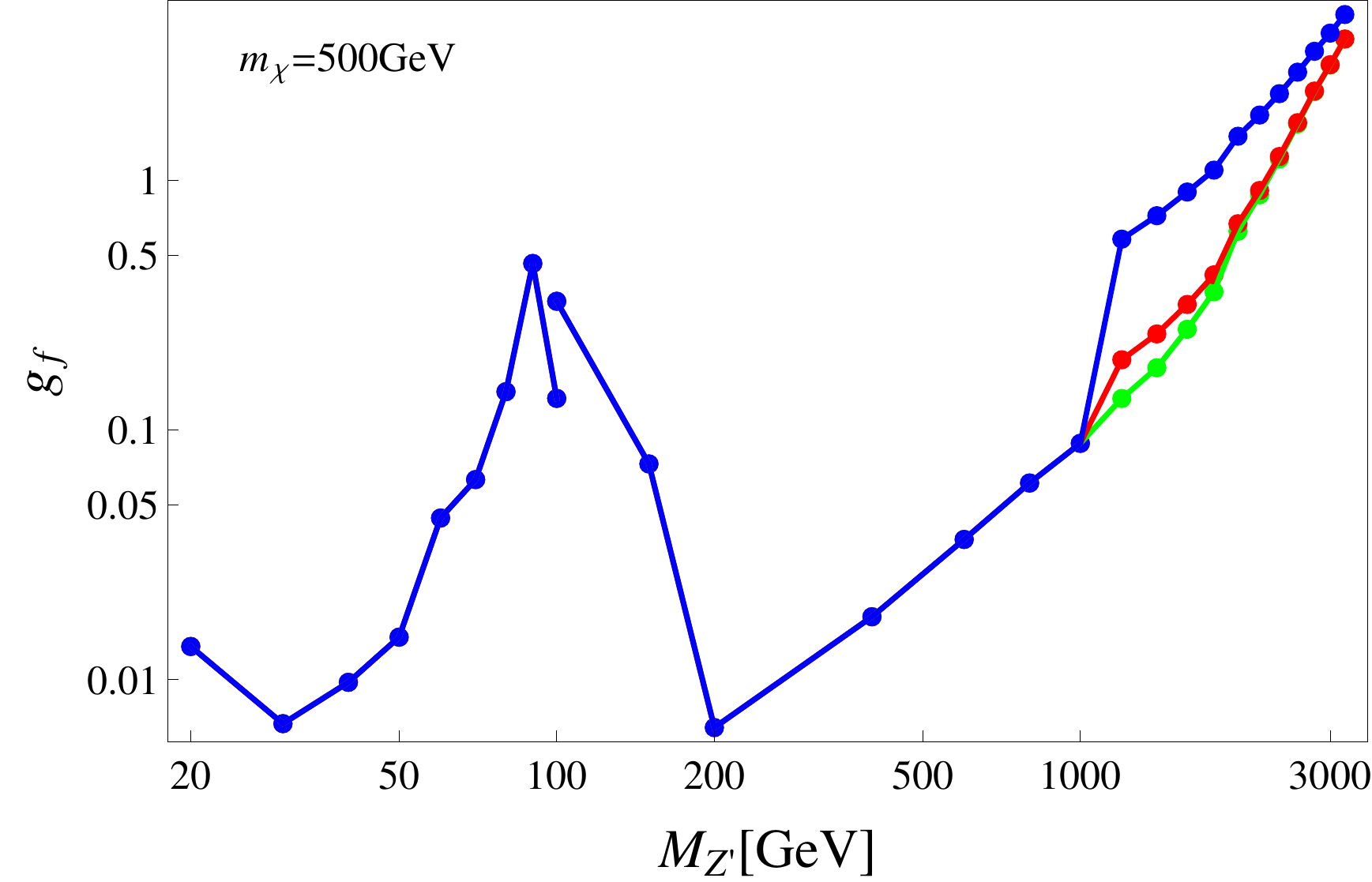}
\includegraphics[scale=0.4]{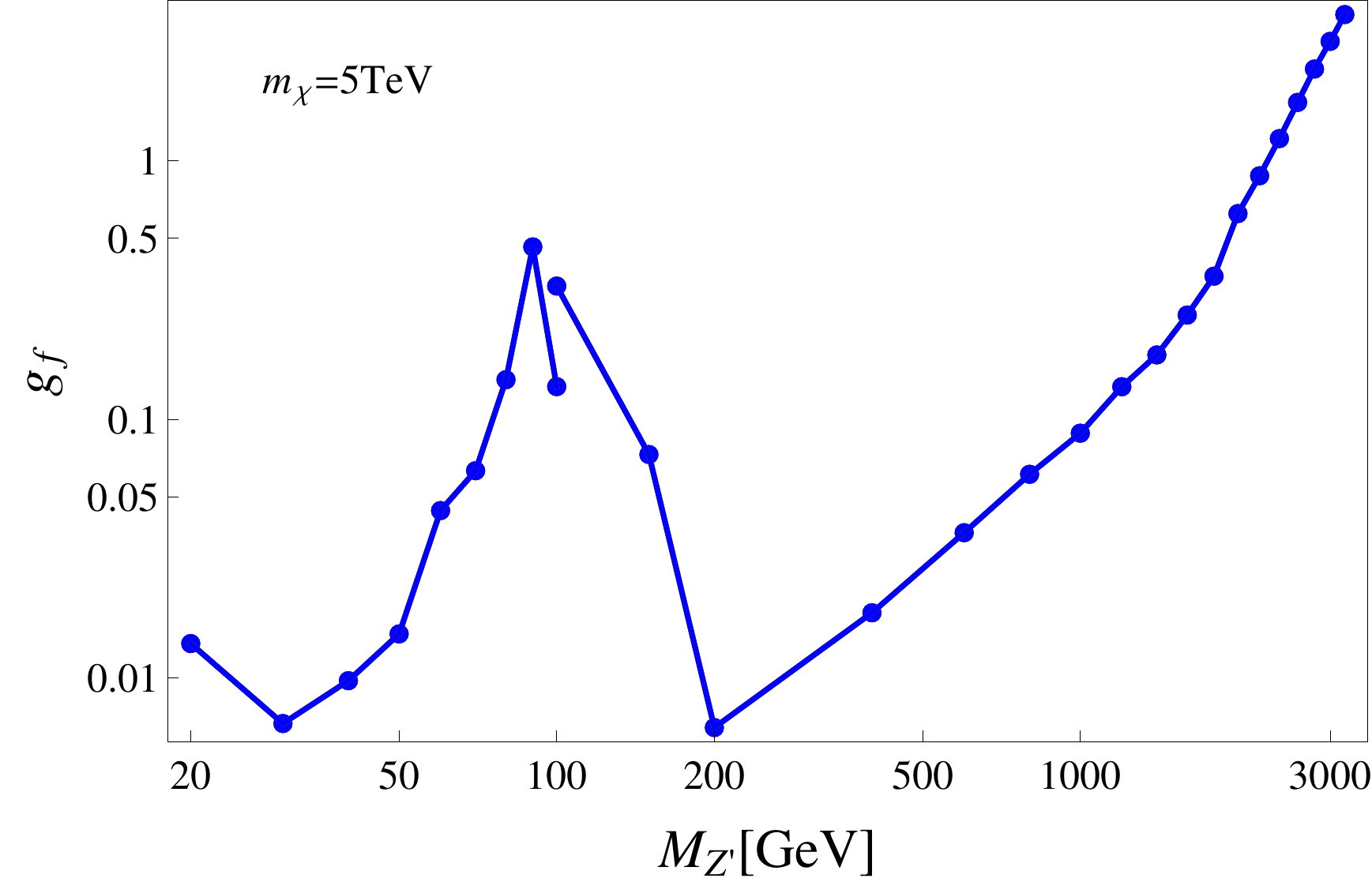}
\caption{The 95\% CL exclusion regions from the searches for dimuon resonances at the 7 and 8 TeV LHC. Four different DM masses and three DM-$\zp$ couplings were chosen to illustrate the collider bounds from those experiments. For $m_\chi\geq 50$ GeV, the constraints for the various $g_\chi$ degenerate into a single bound in the region $\mzp\leq 2m_\chi$. The lines are discontinued at $\mzp=100$ GeV, the point we chose to switch from the CMS 7 TeV data~\cite{Chatrchyan:2013tia} to the ATLAS 8 TeV data~\cite{Aad:2014cka}.}
\label{mzpgf}
\end{figure}

We employ the CL$_\text{S}$ method~\cite{Cowan:2010js} to determine the 95\% confidence limit regions on the $\mzp$ {\it versus} $g_f$ parameter space. First we calculate the related $q$-statistic: $q(\mu_s)=\chi^2(\mu_s)-\chi^2(\hat{\mu}_s)$ if $\mu_s>\hat{\mu}_s$, and 0 otherwise, where $\hat{\mu}_s$ is the best fit for the signal strenght. After that we obtain the bounds by requiring
\begin{equation}
\hbox{CL}_\text{S}=\frac{1-\Phi(\sqrt{q(\mu_s)})}{1-\Phi(\sqrt{q(\mu_s)})-\Phi(\sqrt{q_A(\mu_s)})}=0.05
\label{cls}
\end{equation}
The function $\Phi$ is the cumulative probability function of the standard normal distribution and $q_A(\mu_s)$ is the value of the $q$-statistic calculated assuming $d_i=\hat{\mu}_b b_i$, that is, when data are assumed to be represented by the best background model. Fixing the DM mass and its coupling $g_\chi$ to the $\zp$ boson, we seek for the solution to Eq.~(\ref{cls}) in the ($\mzp$, $g_f$) plane as shown in Fig.~(\ref{mzpgf}). 

In the upper left panel we show the $m_\chi=10$ GeV case for three different $g_\chi$ values: the lower green lines for $g_\chi\leq 0.1$, the middle red ones for $g_\chi=1$, and the upper black ones at the boundary of the perturbative regime $g_\chi=4\pi$. The lines are discontinued at $\mzp=100$ GeV. The constraints for the $\mzp<100$ GeV were derived using the low mass region data of~\cite{Chatrchyan:2013tia}, whilst those in high mass region $\mzp\geq 100$ GeV with data from~\cite{Aad:2014cka}. First, we observe that the excluded regions get larger as $g_\chi$ becomes smaller once the DM cannot compete for decays with leptons and jets as can be seen at the upper row of Fig.~(\ref{brgfmz}). Note that the bounds saturate for $g_\chi< 0.1$.

In the low mass region, the collider constraints are as severe as in high mass region, concerning the values of $g_f$ excluded by the 7 and 8 TeV LHC, respectively, up to $\mzp\sim 50$ GeV. In the $Z$-pole region, the constraints get softened by virtue of the huge SM $Z$ background. Also, for heavier $\zp$ bosons, the production cross sections drop fast and the top decays are turned on rendering the $\sigma(pp\to\zp)\times BR(\zp\to \mu^+\mu^-)$ very small and again escaping the collider constraints.  

As $\chi$ gets heavier, the constraints become increasingly insensitive to the coupling to the $\zp$, once the DM channel remains closed until $\mzp\geq 2m_\chi$. This can be seen in $m_\chi=50$, 500 and 5000 GeV panels of Fig.~(\ref{mzpgf}). For sufficiently heavy DM or with suppressed couplings to $\zp$, couplings between the vector mediator and SM fermions as low as $\sim 5\times 10^{-3}$ are excluded at 95\% CL for $\mzp\sim 30$ and $200$ GeV as we observe in Fig.~(\ref{mzpgf}). These particular masses are a result of the trade off among the size of $\zp$ cross section, the branching ratio to leptons, and the relative distance of the $Z$-pole mass region.

Comparing our 95\% CL limits on $g_f$ with those of Ref.~\cite{Hoenig:2014dsa} for the $Z-Z^\prime$ mixing parameter $\epsilon$, after translating their $g_{ffZ^\prime}$ coupling in terms of our $g_f$, we found agreement in their order of magnitude in the small mass region. The agreement is better for larger $\kappa$ which parametrizes the level of backgrounds systematics in Ref.~\cite{Hoenig:2014dsa}. It should be noted that the mixed $\zp$ model~\cite{Hoenig:2014dsa} assumes vector-axial couplings between $\zp$ and the SM fermions, but it makes a little difference concerning the collider bounds.

\section{Dark Matter Phenomenology}

\noindent
In this section we compare limits from collider searches with the constraints arising from DM phenomenology. These constraints consist in the requirement of the correct DM relic density and the compatibility with limits from both Direct (DD) and Indirect (ID) DM searches. The constraints are individually briefly illustrated below.

\subsection{Relic Density and Indirect Detection}

\noindent
The DM relic density is determined, for the range of couplings considered in our study, by the paradigm of thermal decoupling; as a consequence the experimentally favored value $\Omega h^2 \approx 0.11$ \cite{Ade:2015xua} corresponds to a suitable value of the DM thermally averged pair annihilation cross-section. The DM features two types of annihilation channels. The first is into SM fermions. The corresponding cross-section, originated by s-channel exchange of the $Z^{\prime}$, is given by:

%Note that we deliberately do not include a $\bar{f} \gamma^{\mu} f Z'_{\mu}$ term in order to avoid generating a large vector amplitude for elastic scattering. The annihilation cross section in this case is given by:
%

\begin{eqnarray}
\label{one}
\sigma &=& \sum_{f} \frac{n_c }{12 \pi  \left[ \left(s-m_{Z'} ^2\right)^2  + m_{Z'}^2 \Gamma_{Z'}^2 \Gamma_{Z^{\prime}} \right]} \sqrt{\frac{1-4 m_f^2/s}{1-4 m_{\chi }^2/s}} \\
&\times& g_{f}^2 \Bigg[g_{\chi a}^2 \bigg\{4 m_{\chi }^2 \bigg[m_f^2 \pL 7- \frac{6s}{m_{Z'}^2}+\frac{3s^2}{m_{Z'}^4} \pR-s \bigg]+s  \left(s-4 m_f^2\right)\bigg\} +g_{\chi v}^2  (s-4 m_f^2) (2 m_{\chi}^2+s)  \Bigg], \label{sigsdirV} \nonumber
\end{eqnarray}
where $n_c=3$ (1) for annihilations to quarks (leptons), $\sqrt{s}$ is the center-of-mass energy of the collision, and $\Gamma_{Z'}$ is width of the $Z'$:
\begin{eqnarray}
\Gamma(Z') &=&\sum_f 
\theta(m_{Z'}-2m_f) \frac{n_c m_{Z'} }{24 \pi}  \sqrt{1-\frac{4 m_f^2}{m_{Z'}^2}} \bL g_{f}^2 \left(1-\frac{4 m_f^2}{m_{Z'}^2} \right)+g_{f}^2 \left(1+ 2 \frac{m_f^2}{m_{Z'}^2} \pR \bR \nonumber\\
&& \theta(m_{Z'}-2m_\chi) \frac{ m_{Z'} }{24 \pi}  \sqrt{1-\frac{4 m_\chi^2}{m_{Z'}^2}} \bL g_{\chi a}^2 \left(1-\frac{4 m_\chi^2}{m_{Z'}^2} \right)+g_{\chi v}^2 \left(1+ 2 \frac{m_\chi^2}{m_{Z'}^2} \pR \bR \nonumber\\
\end{eqnarray}
An analytic expression of the thermally averaged cross-section can be obtained through the velocity expansion \cite{Arcadi:2014lta,Alves:2015pea}:
%Although we will use the full expression given in Eq.~\ref{one} for the calculation of the relic abundance, it is illustrative to expand the cross section in powers of velocity.  Presented this way, the cross section (for each SM fermion species) is given by:
\begin{eqnarray}
\label{eq:sigmaff}
\sigma v &\approx& \frac{n_c \sqrt{1-m_f^2/m_\chi^2}}{2 \pi m_{Z'}^4  \left(m_{Z'}^2-4 m_{\chi }^2\right)^2} \,\, g_{f}^2 \Bigg[  m_f^2 g_{\chi a}^2 \left(m_{Z'}^2-4 m_{\chi }^2\right)^2+2 g_{\chi v}^2 m_{Z'}^4 \left(m_{\chi }^2-m_f^2\right) \Bigg] \\
&-& \frac{n_c v^2}{48 \pi  m_{Z'}^4 m_{\chi }^2 \sqrt{1-m_f^2/m_{\chi }^2} \left(4 m_{\chi }^2-m_{Z'}^2\right)^3} 
\,\, g_{f}^2  \Bigg[ g_{\chi a}^2 \left(m_{Z'}^2-4 m_{\chi }^2\right) \times
\nonumber
\\
&&
\Big(m_f^4 \left(-72 m_{Z'}^2 m_{\chi }^2+17 m_{Z'}^4+144 m_{\chi }^4\right) 
+m_f^2 \left(48 m_{Z'}^2 m_{\chi }^4-22 m_{Z'}^4 m_{\chi }^2-96 m_{\chi }^6\right)+8 m_{Z'}^4 m_{\chi }^4\Big) \nonumber\\  
 & - & 2 g_{\chi v}^2 m_{Z'}^4 \left(m_f^2-m_{\chi }^2\right) \Big(4 m_{\chi }^2 \left(m_{Z'}^2-17
   m_f^2\right)+5 m_f^2 m_{Z'}^2+32 m_{\chi }^4\Big) \Bigg] \nonumber.
\end{eqnarray}
In addition, if $m_\chi > m_{Z^{\prime}}$, the t-channel induced $\bar \chi \chi \rightarrow Z^{\prime} Z^{\prime}$ process is kinematically allowed. The analytic expression of $\sigma (s)$ is rather contrived. We will then just report the velocity expansion given by:
\begin{align}
\label{eq:sigmavZZ}
& \langle \sigma v \rangle_{Z^{\prime} Z^{\prime}} \approx \left( \frac{\left(m_\chi^2-m_{Z'}^2\right)^{3/2} \left(g_{a \chi}^4 m_{Z'}^2+2 g_{a \chi}^2 g_{v \chi}^2 \left(4 m_\chi^2-3 m_{Z'}^2\right)+m_{Z'}^2 g_{v \chi}^4\right)}{\pi 
   m_\chi \left(m_{Z'}^3-2 m_\chi^2 m_{Z'}\right)^2}\right.\nonumber\\
&\left.+\frac{\sqrt{m_\chi^2-m_{Z'}^2}}{4 \pi  m_\chi  \left(m_{Z'}^3-2 m_\chi^2 m_{Z'}\right)^4} \left(
m_{Z'}^6 g_{v \chi}^4 \left(76 m_\chi^4+23 m_{Z'}^4-66 m_\chi^2
   m_{Z'}^2\right)\right.\right.\nonumber\\
&\left. \left.-2 g_{a \chi}^2 m_{Z'}^2 g_{v \chi}^2 \left(160 m_\chi^8+21 m_{Z'}^8-182
   m_\chi^2 m_{Z'}^6+508 m_\chi^4 m_{Z'}^4-528 m_\chi^6 m_{Z'}^2\right)\right.\right.\nonumber\\
&\left.\left. g_{a \chi}^4 \left(128 m_\chi^{10}+23 m_{Z'}^{10}-118 m_\chi^2
   m_{Z'}^8+172 m_\chi^4 m_{Z'}^6+32 m_\chi^6 m_{Z'}^4-192 m_\chi^8 m_{Z'}^2\right)\right) \right).
\end{align}

These analytical approximations have been validated by numerically computing the thermally averaged cross-sections through the package Micromegas \cite{Belanger:2014vza}.

Few remarks are in order:

{\color{blue} (i)} Notice that as long as  $g_{\chi v} \ll g_{\chi a}$ the annihilation cross-section into SM fermions is s-wave dominated, with the dark matter annihilating nearly equally to all SM fermions, except for the color index, which makes the overall annihilation to be mostly into quarks;

{\color{blue} (ii)} The term that goes with $g_{\chi v}^2$, not helicity suppressed, gives rise to a detectable indirect detection signal at Telescopes.

{\color{blue} (iii)} The term proportional to $g_{a\chi}$ is velocity suppressed;

{\color{blue} (iv)} When the annihilation into $Z^{\prime}$ pairs is turned on, even the term proportional to $g_{a\chi}$ is no longer velocity suppressed.

{\color{blue} (v)} If we had taken $g_{\chi v}=0$, as would occur for Majorana dark matter, the $Z^{\prime}$ resonance would not have been present, since the pole ($m_{Z^{\prime}}^2-4m_{\chi}^2$) in the numerator cancels out with the denominator.

Keeping that in mind, we have delimited the region that sets the right relic abundance as well as the indirect detection limits from the Fermi-LAT telescope from the observation of dwarf spheroidal galaxies \cite{Fermi-LAT:2016uux} \footnote{See \cite{Hooper:2012sr,Bringmann:2012ez,Bergstrom:2013jra,Gonzalez-Morales:2014eaa,Profumo:2016idl,Caputo:2016ryl} for competitive limits.}.

\begin{figure}[htb]
\begin{center}
\includegraphics[width=7 cm]{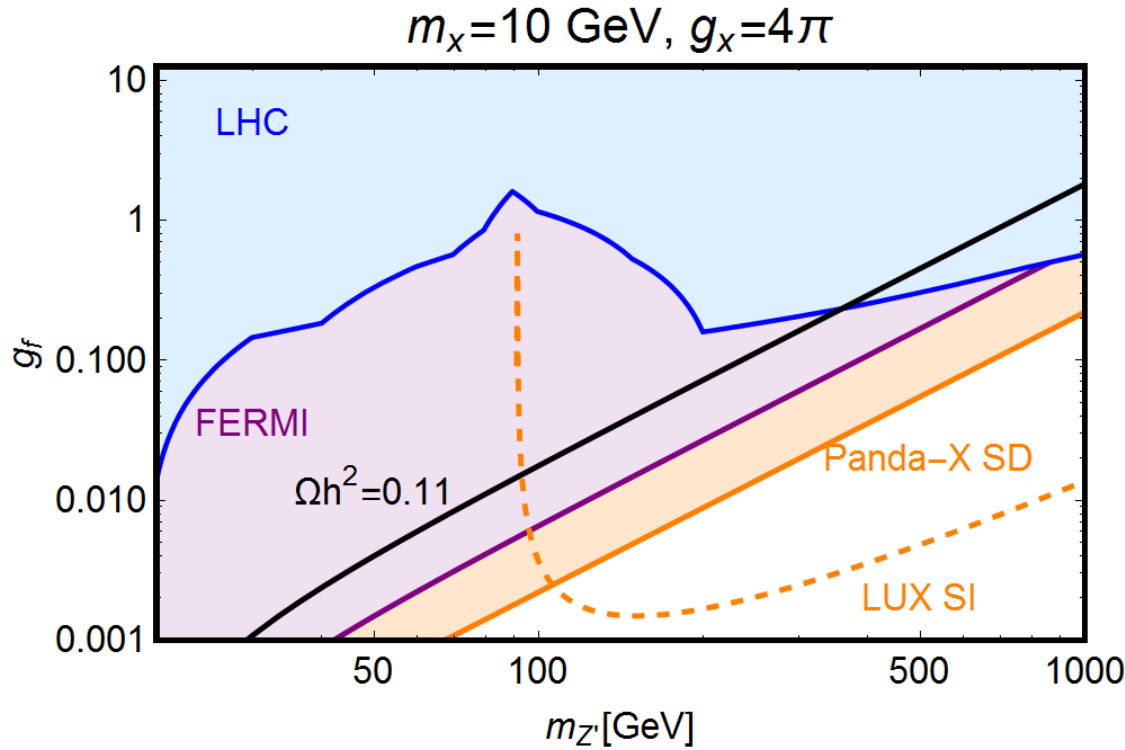}
\includegraphics[width=7 cm]{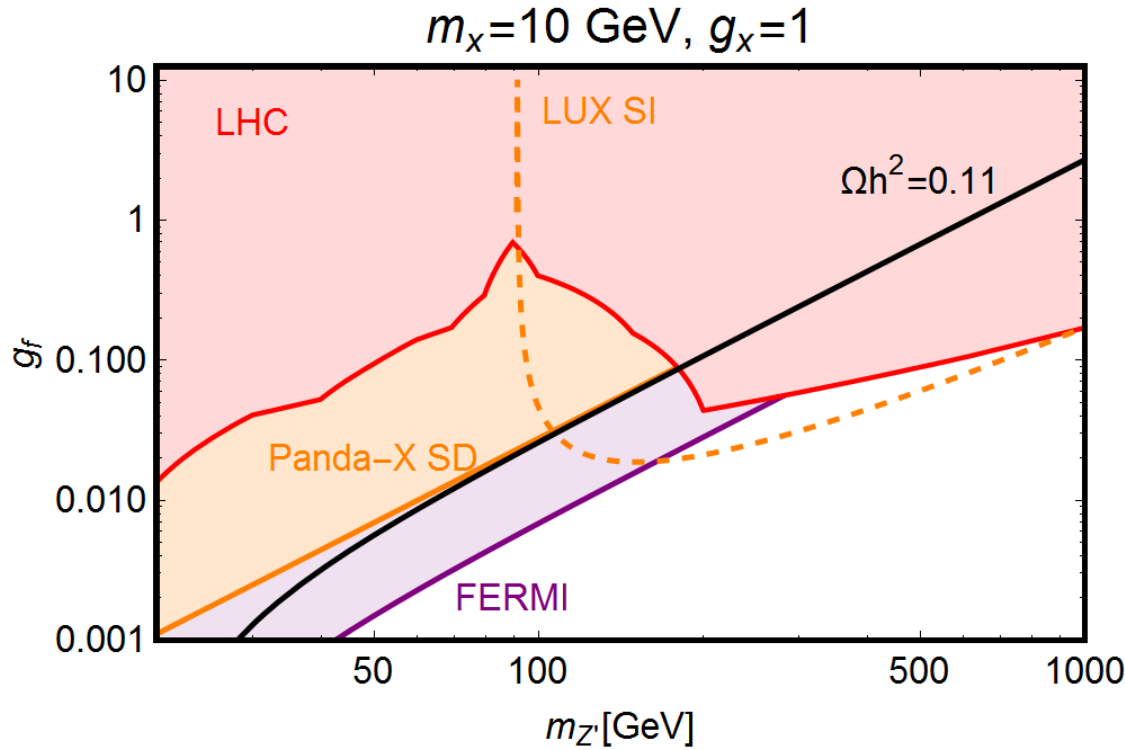}\\
\includegraphics[width=7 cm]{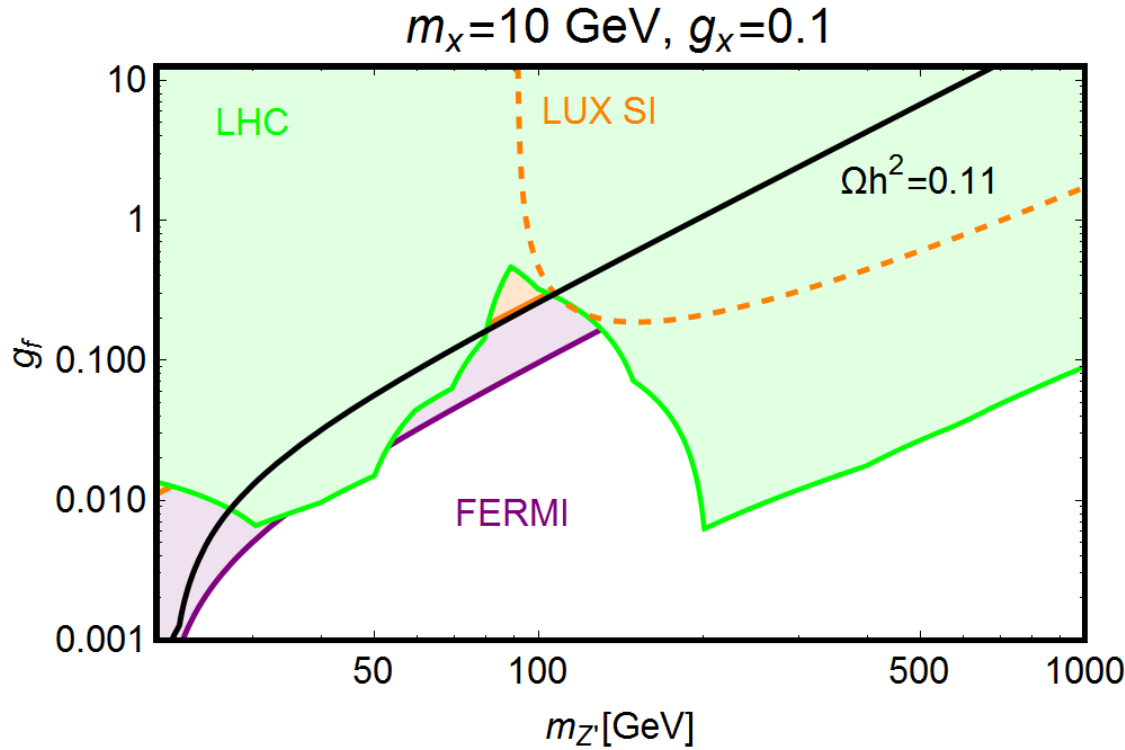}
\end{center}
\caption{{\it Results for} $m_{\chi}=10$~GeV and $g_{\chi}=4\pi,1$ and $0.1$. Combined upper bounds on the model under study, in the bidimensional plane $(m_{Z^{\prime}},g_f)$ for the assignations of the DM mass $m_\chi$ and coupling $g_\chi$ reported in the different panels. The black lines delimit the correct relic density parameter space. The blue, red and green regions are excluded by LHC data. The orange region represents spin-dependent PANDA-X exclusion region, whereas the dashed curve the spin-independent LUX limit, while in purple FERMI-LAT bound. }
\label{fig:DMplots1}
\end{figure}

\subsection{Direct Dark Matter Detection}

In the case of of a $Z'$ with purely axial couplings to quarks one would expect only the spin-dependent interaction between DM and nucleons to be sizable. These are induced by the combination of the axial couplings of the $Z^{\prime}$ with DM and light quarks and the corresponding cross-section is given by (we will consider only the case of scattering on neutrons since it suffers at the moment the most stringent constraints. Notice that in the case of flavor universal couplings the scattering cross sections on protons and neutrons are substantially equal.):
\begin{align}
\label{equation: SD}
\sigma^\text{SD (per neutron)} &\approx \frac{3 \mu^2_{\chi \text{neut}}}{\pi} \frac{g_{\chi a}^2}{m_{\zp}^4} \Big[ g_{u a}  \Delta_u^\text{neut} +  g_{d a} \left( \Delta_d^\text{neut} + \Delta_s^\text{neut}  \right) \Big]^2
~,
\end{align}
where $g_{u a},g_{d a}$ are the vector-axial couplings between the Z' and the up and down quarks respetively, which we assume to be $g_f$ according to Eq.\ref{eq:Diracfermion}, $\mu_{\chi n}$ is the WIMP-nucleon reduced mass while $\Delta_q^\text{neut}$ are the quark spin fractions of the neutron. We will take these to be $\Delta_u^\text{neut} = -0.42$, $\Delta_d^\text{neut} = 0.85$, $\Delta_s^\text{neut} = -0.08$~\cite{Cheng:2012qr}.

\noindent

The vectorial coupling between the dark matter fermion and the $Z^{\prime}$, $g_{\chi v}$, is completely irrelevant for the spin-dependent scattering as one can see in Eq.\ref{equation: SD}. Although, this coupling even if negligible in the initial Lagrangian, Eq.\ref{eq:Diracfermion}, will be non zero, at the typical energy scales of the scattering processes since they are generated through by computing the renormalization group equations (RGE) as shown in Ref.\cite{DEramo:2016gos} so that a spin-independent cross section is actually induced with, 
\begin{align}
\label{equation: SI}
\sigma^\text{SI (per nucleon)} &\approx \frac{a^2 \mu^2_{\chi n}}{\pi} \Big[ \frac{Z f_\text{prot} + (A-Z) f_\text{neut}}{A} \Big]^2
\nonumber \\
f_\text{prot} &\equiv \frac{g_{\chi \text{v}}}{m_{\zp}^2} \left( 2\tilde{g}_{u \text{v}} + \tilde{g}_{d \text{v}} \right)
\nonumber \\
f_\text{neut} &\equiv \frac{g_{\chi \text{v}}}{m_{\zp}^2} \left( \tilde{g}_{u \text{v}} + 2 \tilde{g}_{d \text{v}} \right)
\end{align}where $g_{uv},g_{dv}$ are the vector couplings between the Z' and the up and down quarks respetively, which we are computed through RGE effects. 

Because of the coherent scattering produced by spin-independent WIMP-nucleon interaction, the spin-independent limits are much more restrictive than the spin-dependent ones, for this reason, the spin-independent scattering even if radiatively induced may provide stronger limits in certain regions of the parameter space we we will show below. For the RGE induced $\tilde{g}_{u,d \text{v}}=\tilde{g}_{u,d \text{v}}(\mu_N),\mu_N \sim 1 ~\mbox{GeV}$ couplings we have adopted, for simplicity, the analytical approximation provided in appendix B of \cite{DEramo:2016gos}, retaining only the dominant contribution, induced by top quark loops, present only above the EW scale, i.e. $m_{Z^{\prime}} \gtrsim m_Z$. For $m_{Z^{\prime}} < m_Z$ the spin-dependent limits from PANDA-X are more restrictive and for this reason the spin-independent ones below the Z-pole are not shown in the figures. In the figures we have considered the most recent limits from spin-dependent limits from the PANDA-X experiment \cite{Fu:2016ega}, spin-independent from LUX \cite{Akerib:2016vxi}.

Note that had we started with a Majorana dark matter particle from the beginning, $g_{v\chi}$ would always have vanished, and the RG running effect would have been irrelevant. In this case, only spin-dependent limits would be applicable, the dark matter relic density annihilation cross section would not significantly change, as well as the collider bounds agreeing with \cite{Escudero:2016kpw}. Altough, we have a sizable
change as far as indirect dark matter detection is concerned since in the case of Majorana (or more in
general only axial couplings of the DM with the Z') DM the s-wave component of the annihilation cross-section is helicity suppressed so at late times the annihilation cross-section of the DM is small. 

That said, our findings are also applicable to Majorana Dark Matter, with mild quantitative changes, by simply ignoring the Fermi-LAT limits, as well as the spin-independent limits arising from the RG running and keeping the PANDA-X spin-dependent bounds. At the end, the model would be less constrained by data, since the spin-independent limits from LUX rule out a significant region of the parameter space.

\begin{figure}[htb]
\begin{center}
\includegraphics[width=7 cm]{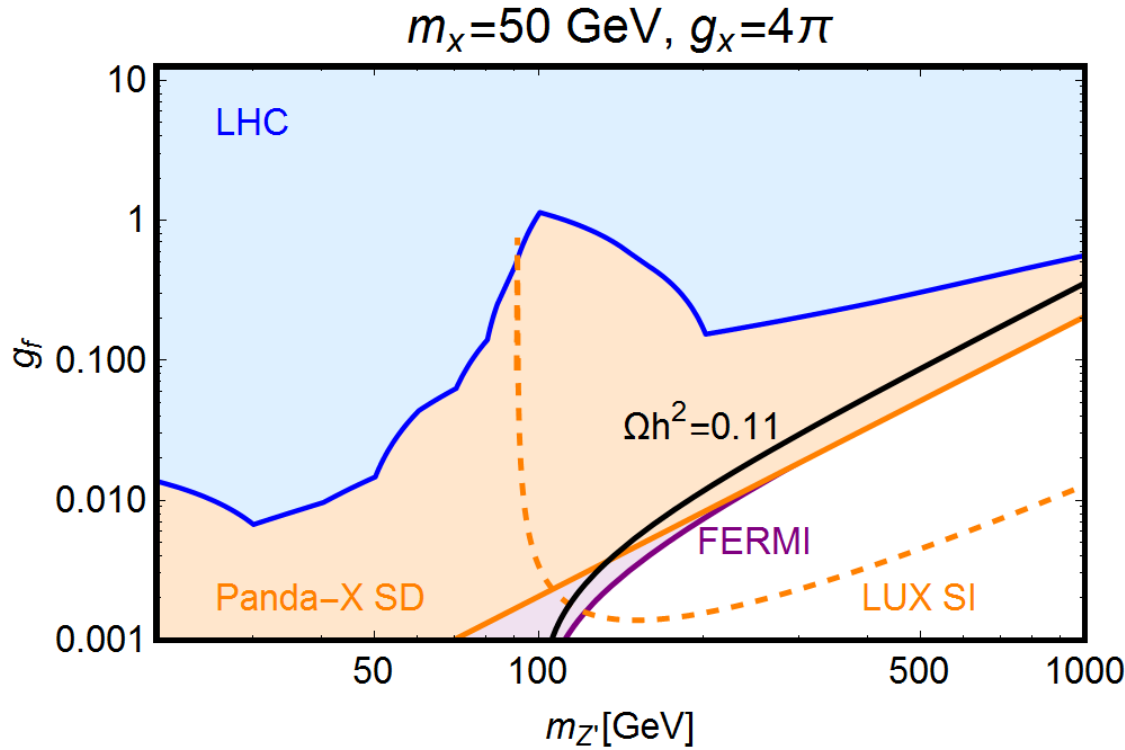}
\includegraphics[width=7 cm]{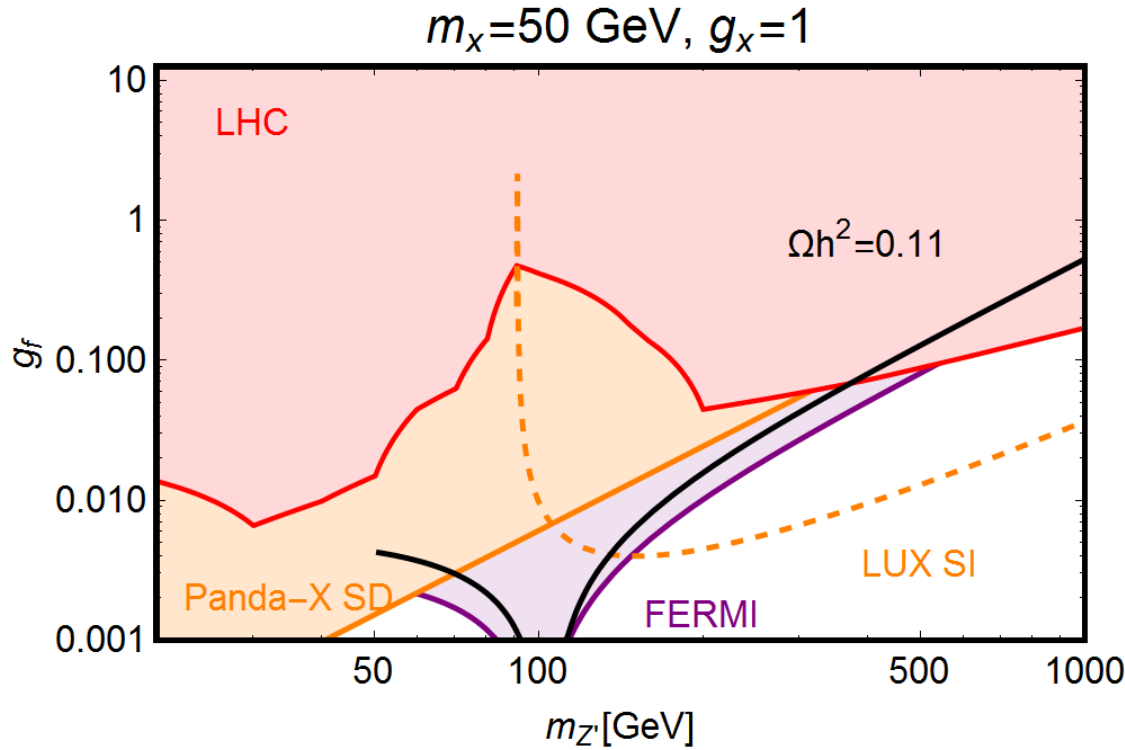}\\
\includegraphics[width=7 cm]{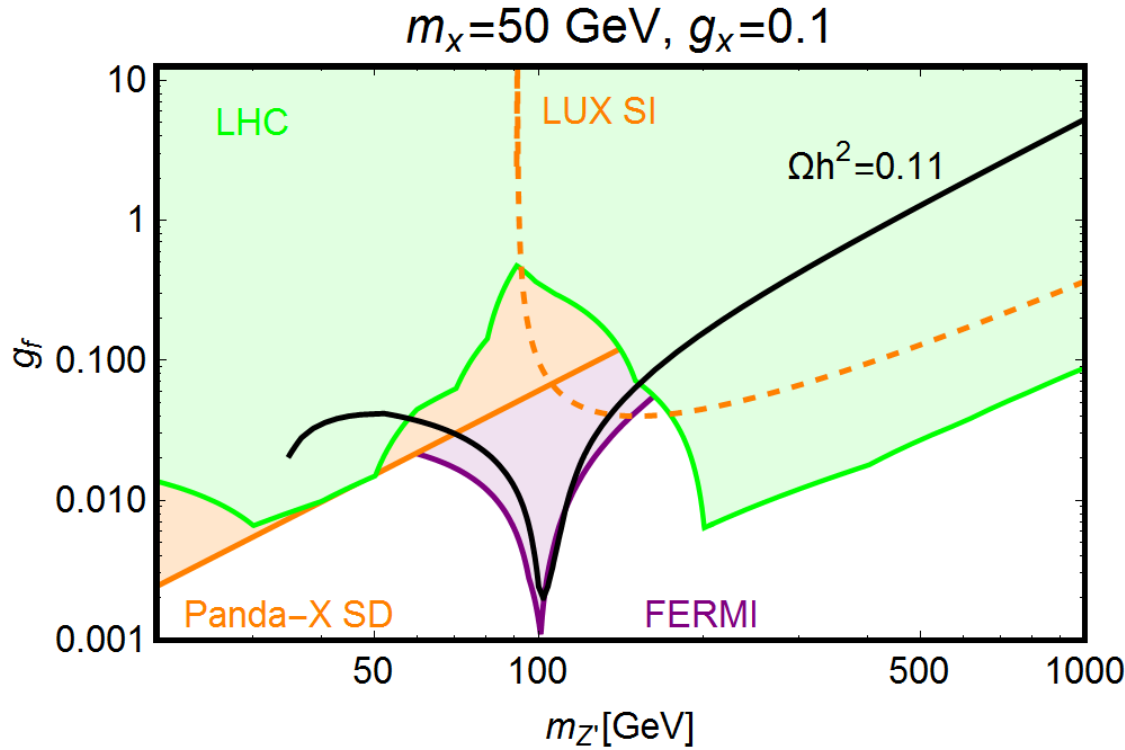}
\end{center}
\caption{{\it Results for} $m_{\chi}=50$~GeV and $g_{\chi}=4\pi,1$ and $0.1$. Combined upper bounds on the model under study, in the bidimensional plane $(m_{Z^{\prime}},g_f)$ for the assignations of the DM mass $m_\chi$ and coupling $g_\chi$ reported in the different panels. The black lines delimit the correct relic density parameter space. The blue, red and green regions are excluded by LHC data. The orange region represents spin-dependent PANDA-X exclusion region, whereas the dashed curve the spin-independent LUX limit, while in purple FERMI-LAT bound.  }
\label{fig:DMplots2}
\end{figure}

\subsection{Summary of results}

\begin{figure}[htb]
\begin{center}
\includegraphics[width=7 cm]{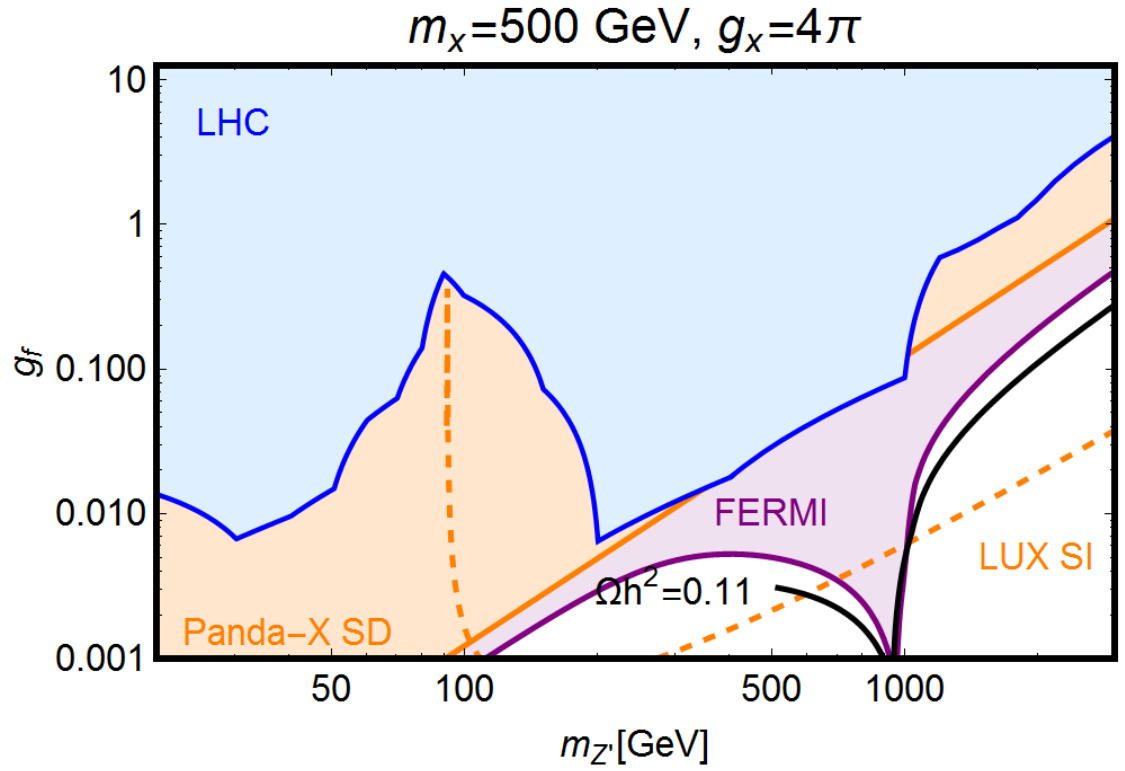}
\includegraphics[width=7 cm]{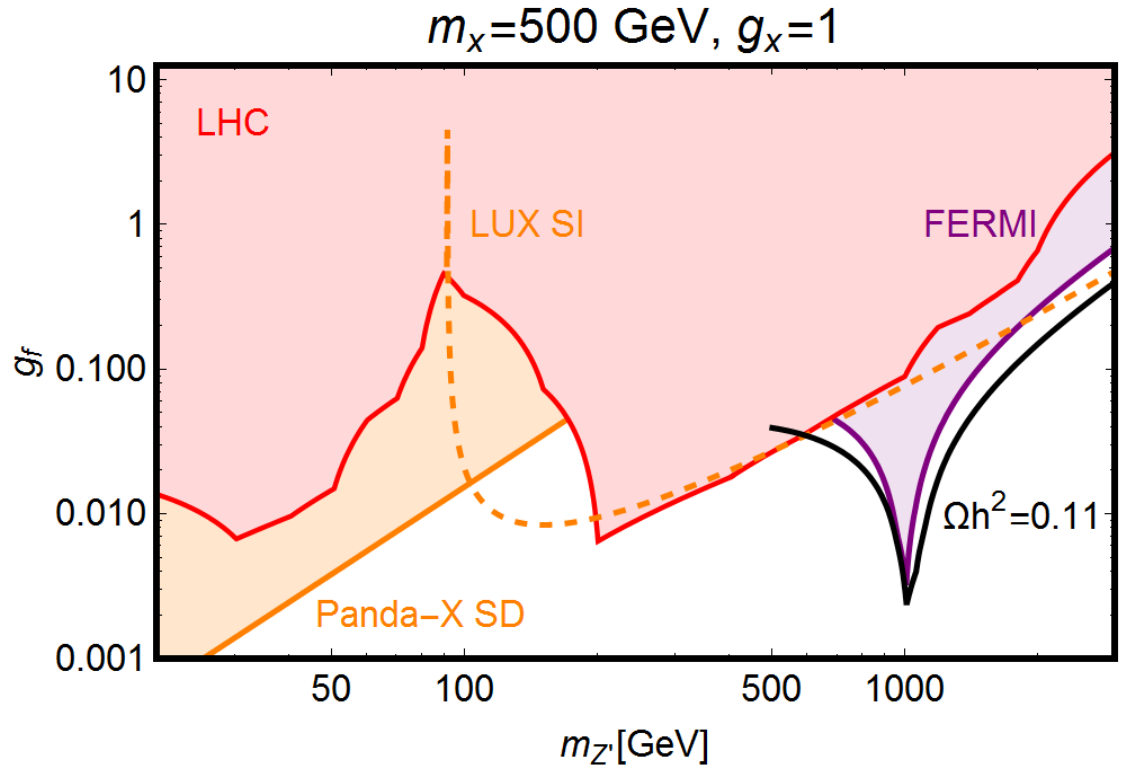}\\
\includegraphics[width=7 cm]{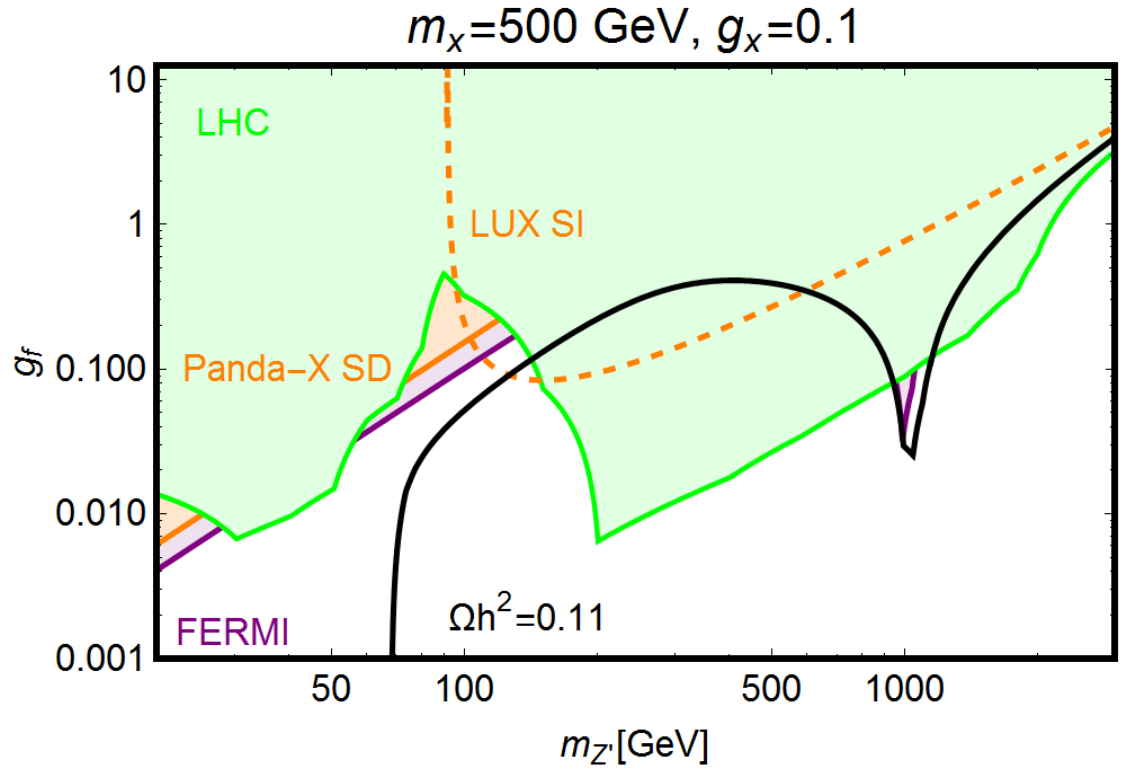}
\end{center}
\caption{{\it Results for} $m_{\chi}=500$~GeV and $g_{\chi}=4\pi,1$ and $0.1$. Combined upper bounds on the model under study, in the plane $(m_{Z^{\prime}},g_f)$ for the a given DM mass $m_\chi$ and coupling $g_\chi$, as reported in the different panels. The black lines delimit the correct relic density parameter space. The blue, red and green regions are excluded by LHC data. The orange region represents spin-dependent PANDA-X exclusion region, whereas the dashed curve the spin-independent LUX limit; finally the purple region indicates the FERMI-LAT bound.   }
\label{fig:DMplots3}
\end{figure}

\noindent
The results of our DM analysis are summarized in Figs. (\ref{fig:DMplots1}-\ref{fig:DMplots3}). Here we have superimposed, for the benchmarks considered in fig. (\ref{mzpgf}), the collider limits from di-muon searches with the isocontours of the correct DM relic density, the limits from spin-dependent cross-section, as recently determined by the PANDA-X experiment \cite{Fu:2016ega}, spin-independent cross-section, as given by LUX \cite{Akerib:2016vxi}, and the most recent limits from indirect searches of DM gamma-ray signals in DSPh \cite{Fermi-LAT:2016uux} \footnote{Low energy observables, such as the muon magnetic moment, also give rise to constraints on the $Z^{\prime}$ mass, but these lie around 100 GeV for couplings of order one, thus not relevant for our reasoning \cite{Lindner:2016bgg}. Moreover notice that our $Z^{\prime}$ model is not ison-spin violating, otherwise a different set of bounds would be applicable \cite{Yaguna:2016bga}.}.   

As already indicated, despite the radiative origin, SI interaction give stronger constraints with respect to SD ones for certain $\zp$ masses. SD limits provide nevertheless a solid complement, especially at light $Z^{\prime}$ masses. Direct detection limits are competitive, or even stronger that the one from LHC for $g_\chi \gtrsim 1$ while the latter dominate for lower values of the DM couplings. Once the FERMI exclusion limit is taken into account, the light DM benchmark, $m_\chi=10\,\mbox{GeV}$ is completely ruled out for $g_f \leq 10^{-3}$. Thermal DM is still in tension with ID limits for mass of 50 GeV ad exception of the pole region, $m_\chi \sim m_{Z^{\prime}}/2$, where mismatch between the annihilation cross-section at freeze-out and at present times is induced by the so called thermal broadening \cite{Gondolo:1990dk,Griest:1990kh}. 

Viable thermal DM can be obtained, far from the pole region, for higher values of the mass,  e.g. $m_\chi=500\mbox{GeV}$, as considered in the last row of fig. (\ref{fig:DMplots3}). Notice that, with the exception of the case $g_\chi=0.1$, there are no regions with the viable DM relic density for $m_\chi > m_{Z^{\prime}}$. Indeed because of the $m_\chi^2/m_{Z^{\prime}}^2$ enhancement and of the high values of the couplings, the DM acquires a very large annihilation cross-section into $Z^{\prime}$ pair as soon as this channel becomes kinematically accessible, so that its relic density is largely suppressed with respect to the experimental expectations. For this same reason, contrary to fig. (\ref{mzpgf}), there are no plots relative to $m_\chi=5\,\mbox{TeV}$ since, in this case, the DM relic density results always several order of magnitude below the correct value, for the couplings choices.

We stress that our results are also applicable to Majorana dark matter, because had we adopted a Majorana dark matter fermion the vectorial coupling $g_{v\chi}$ would have always been zero, and the RG running effect would have been irrelevant. In this case, only spin-dependent limits would have been applicable, with mild changes to the annihilation cross section and collider bounds. As one can see from the figures, the Majorana dark matter setup has a larger region of parameter space allowed by data, if one takes a more conservative indirect detection limit from Fermi-LAT (as we discuss in the next section). In particular, if Fermi-LAT limits are weakened, for $m_{\chi}=50$~GeV, $g_{\chi}=1$ as displayed in Fig.\ref{fig:DMplots2}, a much larger region of the parameter yielding the right relic abundance would be allowed by data.

\section{Galactic Center Excess}
 
An excess in the GeV range has been observed in the Galactic center using data from the Fermi-LAT satellite \cite{Goodenough:2009gk,Hooper:2010mq,Boyarsky:2010dr,Hooper:2011ti,
Abazajian:2012pn,Hooper:2013rwa,Gordon:2013vta,
Huang:2013pda,Daylan:2014rsa,Abazajian:2014fta,Calore:2014xka,TheFermi-LAT:2015kwa}. There are several possible astrophysical explanations for, or caveats to, this excess. An attractive particle physics solution happens to be through annihilations of $30-60$~GeV WIMPs into quarks with an annihilation cross section of $1-3\times 10^{-26}\,{\rm  cm^3s/s}$ normalized to a dark matter local density of $0.4~\mathrm{ GeV/cm^3}$, i.e. slightly below the canonical value \cite{Daylan:2014rsa}. For the light $Z^{\prime}$ model discussed here, the preferred annihilation final states is mostly to quarks, and at the resonance the annihilation cross section today is in the right ballpark of e.g. the results in \cite{Hooper:2014fda}.  

Thus, the model under consideration here can indeed accommodate the GeV excess. However, current constraints from the observation of Dwarf Galaxies using Fermi-LAT data place stringent limits on the annihilation cross section today into quarks \cite{Ackermann:2015zua}. Without including uncertainties in the dark matter content of dwarf galaxies, the WIMP interpretation for the GeV excess is excluded at face value. However, a recent reassessment of the J-factor from the Fermi-LAT team, taking into account systematic uncertainties in the J-factors, weakens their limits by a factor of 2-3, thus showing that there might be still a bit of room left for the WIMP-annihilation hypothesis \cite{Fermi-LAT:2016uux}. Our model thus offers a possible dark matter interpretation for the GeV excess, as long as a conservative limit from Fermi-LAT observation is considered. \footnote{We decided not to go into the details of the astrophysical uncertainties surrounding the GeV excess itself which might shift the favored region downwards, to smaller annihilation cross section today \cite{Calore:2014xka}.}.  
 
\section{Note}

Before submission of our paper we noted the work in
\cite{Escudero:2016kpw} which partially overlaps with ours, but neither incorporated the spin-independent limits resulting from RG running and indirect detection limits, nor performed a detailed collider phenomenology.

\section{Conclusions}

Dirac fermion dark matter models in the context of heavy vector mediators are  forced to live near the $Z^{\prime}$ resonance due to the a combination of spin-independent and LHC bounds. One may switch off the $Z^{\prime}$-fermions vectorial coupling, however, as indeed occurs in some UV-complete models, and consider light $Z^{\prime}$ masses to  circumvent spin-independent direct detection limits and LHC bounds on heavy resonance searches. 

In this work, we have demonstrated that by including the evolution of the vector coupling between the energy scale of the mediator mass and the nuclear energy scale, this coupling, which becomes non-zero, gives rise to stringent independent limits, and that by properly deriving LHC bounds on vector mediators using the $CL_S$ method, the scenario is still rather constrained by data. 

Considering a variety of data, stemming from spin-independent and spin-dependent direct detection, collider, and indirect detection, we showed that only the parameter space near the $Z^{\prime}$ resonance region survives, and that one could possibly accommodate the gamma-ray excess for $m_{\chi}=50$~GeV. Moreover, we have discussed the applicability of our results to Majorana dark matter models.

\section*{Acknowledgements}
The authors warmly thank Paolo Panci for fruitful correspondence, Carlos Yaguna and Manfred Lindner for discussions. A. Alves acknowledges financial support from CNPq (process 307098/2014-1) and FAPESP (process 2013/22079-8). SP is partly supported by the U.S. Department of Energy grant number DE-SC0010107.
Y.~M. acknowledges partial support the ERC advanced grants Higgs@LHC and MassTeV. 
This research was also supported in part by the Research Executive Agency (REA) of the European Union under the Grant Agreement {\bf PITN-GA2012-316704} (``HiggsTools'') and has received funding from the European Union's Horizon 2020 research and innovation programme under the Marie Sklodowska-Curie grant agreement No 674896.

\bibliographystyle{JHEPfixed}
\bibliography{darkmatter}

\end{document}